\documentclass[onecollarge,final]{svjour2}

\smartqed

\usepackage [dvips]{graphicx}
\usepackage{amssymb}
\usepackage{epsfig}
\usepackage{graphicx}
\usepackage{verbatim}

\usepackage[T1]{fontenc}

\usepackage{psfrag}

\usepackage{amsmath}
\usepackage{enumerate}
\usepackage{graphicx}
\usepackage{mathbbol}
\usepackage{amsfonts}
\newcommand{\lmax}{\lambda_{\rm max}}

\newcommand{\DOS}{\rm DOS}

\begin{document}

%\preprint{}

\title{Near-extreme eigenvalues and the first gap of Hermitian random matrices}% repeat the \author .. \affiliation  etc. as needed


\author{Anthony Perret \and Gr\'egory Schehr}

\institute{A. Perret \at Laboratoire de Physique Th\'eorique et Mod\`eles
  Statistiques, Universit\'e Paris-Sud, B\^at. 100, 91405 Orsay Cedex, France\\ \and G. Schehr \at  Laboratoire de Physique Th\'eorique et Mod\`eles
  Statistiques, Universit\'e Paris-Sud, B\^at. 100, 91405 Orsay Cedex, France}   

%\email{anthony.perret@u-psud.fr}
%\email{gregory.schehr@u-psud.fr}

\date{\today}

\maketitle

\begin{abstract}
We study the phenomenon of ``crowding'' near the largest eigenvalue $\lmax$ of random $N \times N$ matrices belonging to the Gaussian Unitary Ensemble (GUE) of random matrix theory. We focus on two distinct quantities: (i) the density of states (DOS) near $\lmax$, $\rho_{\rm DOS}(r,N)$, which is the average density of eigenvalues located at a distance $r$ from $\lmax$ and (ii) the probability density function of the gap between the first two largest  eigenvalues, $p_{\rm GAP}(r,N)$. In the edge scaling limit where $r = {\cal O}(N^{-1/6})$, which is described by a double scaling limit of a system of unconventional orthogonal polynomials, we show that $\rho_{\rm DOS}(r,N)$ and $p_{\rm GAP}(r,N)$ are characterized by scaling functions which can be expressed in terms of the solution of a Lax pair associated to the Painlev\'e XXXIV equation. This provides an alternative and simpler expression for the gap distribution, which was recently studied by Witte, Bornemann and Forrester in {\it Nonlinearity} {\bf 26}, 1799 (2013). Our expressions allow to obtain precise asymptotic behaviors of these scaling functions both for small and large arguments.      
\end{abstract}

%\tableofcontents

\section{Introduction and summary of main results}

Extreme value statistics (EVS) is currently an active subject of studies in various areas of sciences, and in particular in statistical physics \cite{BM97,DM01,LDM03}. The standard 
question which is usually addressed concerns the fluctuations of the maximum $X_{\max}$ among a collections of $N$ random variables $X_1, \cdots, X_N$. However in many circumstances, it is natural to wonder about the ``crowding'' near the maximum $X_{\max}$: is $X_{\max}$ very far from the others or, on the contrary, are there many others close to it \cite{SM2007} ? This type of question arises for instance naturally in the study of complex and disordered systems where the thermodynamical properties are dominated by the low-lying states, close to the ground state \cite{FH1988a,FH1988b,LDM04,MG13}. Such questions about near extreme events play also an important role in natural sciences \cite{Omo1894,Ver69} or in finance \cite{PWHS10}. They were also studied in the context of sporting events, like in marathon packs \cite{SMR08}. 

A natural way to characterize quantitatively this phenomenon of crowding is to study the full order statistics \cite{DN2003} where one considers not only the 
first maximum $M_{1,N} = X_{\max}$ but also the second $M_{2,N}$, the third $M_{3,N}$, $\cdots$, more generally the $k^{\rm th}$ maximum $M_{k,N}$. 
A set of particularly interesting random variables which are sensitive to the crowding of the extremum are the gaps $d_{k,N} = M_{k,N} - M_{k+1,N}$ between successive maxima. While order (or gap) statistics have been widely studied in the past for independent  and identically distributed (i.i.d.) random variables~\cite{DN2003,Gum58}, there exists very few exact results for strongly correlated random variables. Yet, many problems of statistical physics involve order statistics of strongly correlated variables, as studied for branching Brownian motion \cite{BD2009,BD2011} or signals with $1/f^\alpha$ correlations \cite{MOR2011}. Recent analytical progress was achieved for the gap statistics of random walks, which were shown to display a very rich structure \cite{SM12,MMS13}. Another physically relevant instance of strongly correlated random variables where extreme value questions has attracted much attention is the set of eigenvalues of random matrices \cite{Meh91,For10}. In particular, the Tracy-Widom (TW) distributions \cite{TW94a,TW96} which describe the fluctuations of the largest eigenvalues of random matrices belonging to the Gaussian ensembles (orthogonal, unitary and symplectic) of random matrix theory (RMT) are now cornerstones of the theory of EVS for strongly correlated random variables. For these ensembles, the distribution of the $k^{\rm th}$ eigenvalue  can also be computed explicitly as rather simple generalization of the TW distribution \cite{TW94a,TW96,DT05} (we refer the reader to the Refs. \cite{Gus05,ORou10} for the analysis of these distributions in the large $k$ limit where they become Gaussian). Despite the fact that one can formally write the joint probability density function (PDF) of the $k$ first eigenvalues in terms of Fredholm determinants, it is only recently that the first gap between the first two largest eigenvalues of large Gaussian random matrices has been investigated in detail for GUE in Ref. \cite{WBF13} by Witte, Bornemann and Forrester (referred to as WBF in the following). In particular, they obtained an expression for the PDF of the first gap in terms of the components of a solution of a particular isomonodromic problem relating to the Painlev\'e II equation (see Appendix \ref{formula_WBF}). We will show below that the approach developed in this paper allows to obtain an alternative and somewhat simpler expression for this PDF of the first gap in the GUE ensemble.

\begin{figure}
\begin{center}
\includegraphics[width=0.7\linewidth]{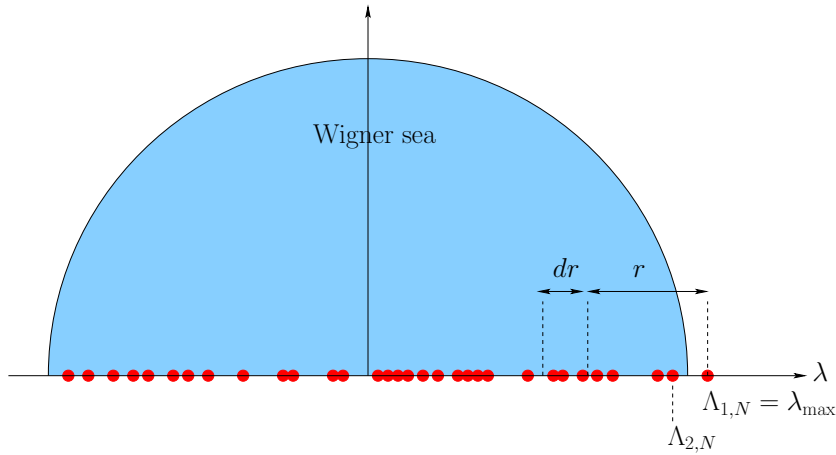}
\caption{Different quantities characterizing the ``crowding'' near the largest eigenvalue $\lmax$ studied in this paper: (i) the mean density of states $\rho_{\rm DOS}(r,N)$ such that $\rho_{\rm DOS}(r,N) dr$ is the mean number of eigenvalues located in the interval $[\lmax-r-dr,\lmax-r]$ and (ii) the PDF $p_{\rm GAP}(r,N)$ of the spacing between the two largest eigenvalues, $p_{\rm GAP}(r,N) dr = {\Pr.} [(\Lambda_{1,N} - \Lambda_{2,N}) \in [r, r+dr]]$.}\label{fig_illust}
\end{center}
\end{figure}\hfill

Another natural quantity to characterize the phenomenon of crowding near the maximum is the (average) density of states (DOS) $\rho_{\rm DOS}(r,N)$
near the maximum, which counts the mean number of random variables $X_i$'s located at a given distance $r$ from the maximum $X_{\max}$~\cite{SM2007}:
\begin{eqnarray}\label{eq:def_DOS_gen}
\rho_{\rm DOS}(r,N) = \frac{1}{N-1} \sum_{i \neq {\rm i_{\max}}} \langle \delta(X_{\max}-X_i -r) \rangle \;,
\end{eqnarray} 
where $X_{\rm i_{\max}} = X_{\max}$ and $\langle \cdots \rangle$ denotes an average over the different realizations of the random variables $X_i$'s. This quantity $\rho_{\rm DOS}(r,N)$ has been studied in mathematical statistics \cite{PS97,PL98} and, more recently, in physics \cite{SM2007}, for i.i.d. and weakly correlated random variables. In particular, it was shown that, depending on whether the tail of the parent distribution of the $X_i$'s decays slower than, faster than, or as a pure exponential, the limiting mean DOS converges to three different limiting forms. More recently, the DOS was computed exactly for Brownian motion \cite{PCMS13} which is thus one rare instance of strongly correlated random variables where this phenomenon of crowding near extremes has been studied analytically. 

Here we will focus on the set of eigenvalues of random matrices belonging to the Gaussian ensembles of RMT. Although the density of near extreme eigenvalues is a natural object to study, it has not -- to our knowledge -- been discussed before in the literature. Here
we consider random Hermitian matrices, of size $N \times N$, belonging to the Gaussian Unitary Ensemble (GUE) of random matrices. The joint PDF of the eigenvalues is given by
\begin{eqnarray}\label{jPDF}
P_{\rm joint}(\lambda_1,\lambda_2,...,\lambda_N)=\frac{1}{Z_N}\prod_{i<j}(\lambda_i-\lambda_j)^2 \, e^{-\sum_{i=1}^{N}\lambda_i^2} \;,
\end{eqnarray}
where the normalization constant is $Z_N = 2^{-\frac{N^2}{2}}(2 \pi)^{\frac{N}{2}} \prod_{j=1}^N j!$. It is well known that the fluctuations of the eigenvalues are characterized by two different scales depending on their location in the spectrum: (i) in the bulk for $\lambda_i/\sqrt{N} = {\cal O}(1)$ and $|\lambda_i| < \sqrt{2N}$ and (ii) at the edge where $|\lambda_i \pm \sqrt{2N}| = {\cal O}(N^{-1/6})$. The existence of these two scales manifests itself in various observables associated to the eigenvalues of GUE (\ref{jPDF}), including their mean density defined as:
\begin{eqnarray}\label{def_rho}
\rho(\lambda,N) = \frac{1}{N} \sum^N_{i=1} \langle \delta(\lambda_i - \lambda) \rangle \;.
\end{eqnarray} 
One has obviously $\rho(\lambda,N) = \rho(-\lambda,N)$ and one can further show that, for large $N$, it exhibits two distinct regimes (for $\lambda > 0$) \cite{Meh91,For10,For93}
\begin{eqnarray}\label{scaling_density}
\rho(\lambda,N) \sim
\begin{cases}
\dfrac{1}{\sqrt{N}} \rho_{\rm bulk}\left(\dfrac{\lambda}{\sqrt{N}} \right) \;, \; &\lambda = {\cal O}(\sqrt{N}) \; \& \; \lambda < \sqrt{2N} \;, \\
\\
\sqrt{2} N^{-5/6} \rho_{\rm edge}\left( (\lambda-\sqrt{2N})\sqrt{2}N^{1/6} \right) \;, \; &|\lambda - \sqrt{2N}| = {\cal O}(N^{-1/6}) \;.
\end{cases}
\end{eqnarray}
In Eq. (\ref{scaling_density}), $\rho_{\rm bulk}(x)$ is the Wigner semi-circle \cite{Meh91,For10}:
\begin{eqnarray}\label{eq:semi_circle}
\rho_{\rm bulk}(x) = \rho_W(x) = \frac{1}{\pi} \sqrt{2-x^2} \;,
\end{eqnarray}
while $\rho_{\rm edge}(x)$ is given by the Airy kernel at coinciding point \cite{For93} (see also Ref. \cite{BB91}),
\begin{eqnarray}
\rho_{\rm edge}(x) = [{\rm Ai}'(x)]^2 - x {\rm Ai}^2(x) \;,
\end{eqnarray}
whose asymptotic behaviors are given by \footnote{note that we have corrected a typo appearing in the large $x$ asymptotic behavior of $\rho_{\rm edge}(x)$ given in Eq. (3.11b) of Ref.~\cite{For93}.}
\begin{eqnarray}\label{asympt_edge}
\rho_{\rm edge}(x) \sim
\begin{cases}
&\dfrac{1}{\pi}\sqrt{-x} \;, \; x \to -\infty \;,\\
& \\
& \dfrac{1}{8 \pi x} e^{-\frac{4}{3}x^{3/2}}\;, \; x \to \infty \;.
\end{cases}
\end{eqnarray}
Interestingly, one can check that these two regimes for $\rho(\lambda,N)$, the ``bulk'' one and the ``edge'' one in Eq. (\ref{scaling_density}), 
perfectly match when $\lambda$ approaches the value $\sqrt{2N}$ from below. Indeed, when $\lambda \to \sqrt{2N}$ from below, $\rho(\lambda,N)$ can be replaced by the Wigner semi-circle (\ref{eq:semi_circle}), which gives:  
\begin{eqnarray}\label{eq:matching_left}
\rho(\lambda,N) \sim \frac{2^{3/4}}{\pi} N^{-3/4} \left({\sqrt{2N} - \lambda}\right)^{1/2} \;, \; \lambda \to \sqrt{2N}^{\,-} \;.
\end{eqnarray}
This behavior (\ref{eq:matching_left}) coincides with the left tail of the scaling function $\rho_{\rm edge}(x)$ in Eq. (\ref{asympt_edge}). Indeed, when the deviation from $\sqrt{2N}$ is large, $\sqrt{2N} - \lambda \sim {\cal O}(\sqrt{N})$, we can substitute in the second line of Eq. (\ref{scaling_density}) the left tail asymptotic behavior of $\rho_{\rm edge}(x)$ in (\ref{asympt_edge}), which gives
\begin{eqnarray}\label{eq:matching_right}
\rho(\lambda,N) \sim \sqrt{2}N^{-5/6} \frac{1}{\pi} \left(\sqrt{2}N^{1/6}(\sqrt{2N}-\lambda) \right)^{1/2} \;, \; \lambda \to \sqrt{2N}^{\,-} \;,
\end{eqnarray} 
which after a trivial rearrangement coincides with Eq. (\ref{eq:matching_left}). Note also that the right tail of $\rho_{\rm edge}(x)$ in Eq. (\ref{asympt_edge}) matches, as it should, with the right tail of the TW distribution for GUE \cite{For10} (see also \cite{MSVV13}). 

\begin{figure}
\begin{center}
\psfrag{sqrt(n)rho}[c][][2.5]{$\sqrt{N}\rho(r,N)$}
\psfrag{sqrt(n)Tmax}[c][][2.5]{$\sqrt{N}\rho_{\rm DOS}(r,N)$}
\psfrag{sqrt(2)}[c][][2]{$\sqrt{2}$}
\psfrag{-sqrt(2)}[c][][2]{$-\sqrt{2}$}
\psfrag{2sqrt(2)}[c][][2]{$2\sqrt{2}$}
\psfrag{0}[c][][2]{$0$}
\psfrag{0.5}[c][][2]{$0.5$}
\psfrag{r/sqrt(n)}[c][][2.5]{$r N^{-1/2}$}
\rotatebox{-90}{\resizebox{80mm}{130mm}{\includegraphics{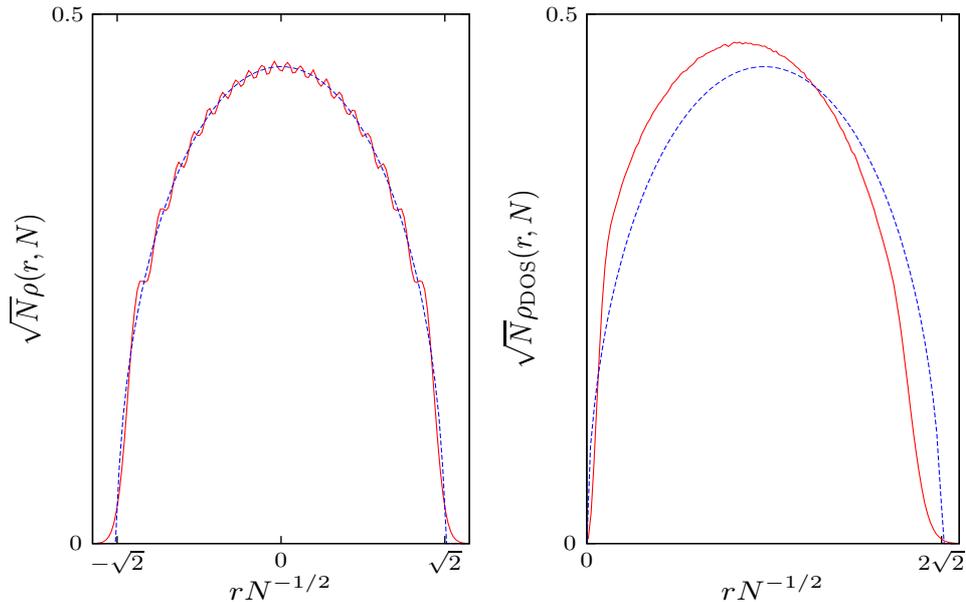}}}
\caption{Plots of the density of eigenvalues $\rho(r,N)$ and $\rho_{\rm DOS}(r,N)$, evaluated numerically for GUE matrices of sizes $N=20$ (red solid curve) and the two asymptotic bulk behavior (blue dashed curve).}
\label{fig_illust_DOS}
\end{center}
\end{figure}

Here we focus on the density of eigenvalues near the largest one, $\lmax = \max_{1\leq i \leq N} \lambda_i$ and consider, as in Eq. (\ref{eq:def_DOS_gen}), the quantity
$\rho_{\rm DOS}(r,N)$ defined as (see Figs. \ref{fig_illust} and \ref{fig_illust_DOS}) 
\begin{eqnarray}\label{eq:def_rho_eigen}
\rho_{\rm DOS}(r,N)=\frac{1}{N-1} \sum_{\underset{i \ne i_{\rm max}}{i=1}}^{N} \langle \delta(\lambda_{\rm max}-\lambda_i-r) \rangle \;,
\end{eqnarray}
where $i_{\rm max}$ is such that $\lambda_{i_{\max}} = \lambda_{\rm max}$ and 
$\langle \cdots \rangle$ means an average taken with the weight in (\ref{jPDF}). It is normalized according to
\begin{eqnarray}
\int_0^{\infty}{\rm d}r\, \rho_{\rm DOS}(r,N)=1 \;.
\end{eqnarray}
In this paper, we show how to compute this density $\rho_{\rm DOS}(r,N)$, exactly for all $N$, in terms of an unconventional family of orthogonal polynomials (OP), which were introduced in Ref. \cite{NM11} to provide a simple derivation of the Tracy-Widom distribution for GUE. Later on they were used in the context of multi critical matrix models in Ref. \cite{AA13,AZ13}. They are unconventional in the sense that they are defined on a semi-infinite real interval (in the mathematical literature they are known to arise in 
the study of Janossy densities for unitary matrix ensembles \cite{CK08,BS03}). These OP are 
monic polynomials $\pi_k(\lambda,y)$ (which are polynomials of degree $k$ of the variable $\lambda$, while $y$ is a parameter) defined by \cite{NM11}
\begin{eqnarray}\label{def_op_intro}
\left\{
     \begin{array}{lr}
			\langle \pi_k, \pi_{k'}\rangle =  \int_{-\infty}^{y} {\rm d}\lambda \, \pi_k(\lambda,y) \pi_{k'}(\lambda,y) e^{-\lambda^2} = \delta_{k,k'}h_k(y)\\
			\\
			\pi_k(\lambda,y)=\lambda^k+...
     \end{array}
\right.
\end{eqnarray}
By performing a large $N$ analysis we show that for large $N$, $\rho_{\rm DOS}(r,N)$ exhibits two different scaling regimes: (i) the bulk regime, where $r = {\cal O}(\sqrt{N})$ and (ii) the edge regime where $r = {\cal O}(N^{-1/6})$, which 
necessitates to study the aforementioned system of OP (\ref{def_op_intro}) in a double scaling limit which is made precise below. Our main results for $\rho_{\rm DOS}(r,N)$ for large $N$ can thus be summarized as follows
\begin{eqnarray}\label{DOS_main}
\rho_{\rm DOS}(r,N) \sim
\begin{cases}
\dfrac{1}{\sqrt{N}} \tilde \rho_{\rm bulk} \left( \dfrac{r}{\sqrt{N}} \right) \;, &r = {\cal O}(\sqrt{N}) \; \& \; 0< r < 2\sqrt{2N} \;, \\
& \\
\sqrt{2} N^{-5/6} \tilde \rho_{\rm edge}\left( r\sqrt{2}N^{1/6} \right) \;, \; &r = {\cal O}(N^{-1/6}) \;,
\end{cases}
\end{eqnarray}
where $\tilde \rho_{\rm bulk}(x)$ and $\tilde \rho_{\rm edge}(\tilde r)$ are scaling functions which we compute exactly. The first one 
is simply the Wigner semi-circle law with a shifted argument (see also Fig. \ref{fig_illust_DOS})
\begin{eqnarray}\label{eq:tilde_bulk}
\tilde \rho_{\rm bulk}(x) = \rho_W(\sqrt{2}-x)=\frac{1}{\pi} \sqrt{x(2\sqrt{2}-x)} \;.
\end{eqnarray}
It can be simply understood from the definition of $\rho_{\rm DOS}(r,N)$ in (\ref{eq:def_rho_eigen}) as for large $N$, $\lambda_{\max} \sim \sqrt{2N}$ while its fluctuations are of order ${\cal O}(N^{-1/6})$ and described by the TW distribution. Hence, if $r =Ê{\cal O}(\sqrt{N})$ in (\ref{eq:def_rho_eigen}) it is insensitive to the fluctuations of $\lmax$ and its PDF can be simply replaced in (\ref{eq:def_rho_eigen}) by a delta function $\delta(\lmax - \sqrt{2N})$. It thus follows that $\rho_{\rm DOS}(r,N) \approx \rho(\sqrt{2N} - r,N)$ which using (\ref{eq:semi_circle}) yields the first line of (\ref{DOS_main}) together with the expression in (\ref{eq:tilde_bulk}). 

The expression of the scaling function $\tilde \rho_{\rm edge}(x)$ has a more complicated and interesting analytical structure. We show that it can be written as 
\begin{eqnarray}\label{rho_intro_simplif}
\tilde \rho_{\rm edge}(\tilde r) = \frac{2^{1/3}}{\pi} \int_{-\infty}^\infty \left[\tilde f^2(\tilde r,x) - \left( \int_x^\infty q(u) \tilde f(\tilde r,u) du  \right)^2 \right] {\cal F}_2(x) \, dx \;,
\end{eqnarray}
where ${\cal F}_2(u)$ is the TW distribution associated to GUE. We recall that it can be written as \cite{TW94a}
\begin{eqnarray}\label{def_TW2}
{\cal F}_2(x) = \exp{\left[-\int_x^\infty (u-x) q^2(u) du\right]} \;,
\end{eqnarray}
where $q(x)$ in Eqs.  (\ref{rho_intro_simplif}) and (\ref{def_TW2}) is the Hastings-McLeod solution of Painlev\'e II with parameter $\alpha = 0$~\footnote{The Painlev\'e II equation with parameter $\alpha$ reads $q_{\alpha}''=2q_{\alpha}^3+xq_{\alpha}-\alpha$, see also below in Eq. (\ref{P2_alpha1/2}). The Hastings-McLeod solution has the asymptotic behavior $q_{\alpha}(x) \sim \alpha/x$, for $x \to \infty$.}
\begin{eqnarray}\label{hasting}
\left\{
     \begin{array}{lr}
			q''=2q^3+x \, q \;,\\
			\\
			q(x)\sim {\rm Ai}(x) \textrm{ for } x \to \infty \;,
     \end{array}
\right.
\end{eqnarray}
with $q \equiv q(x)$ and where ${\rm Ai}(x)$ is the Airy function. In Eq. (\ref{rho_intro_simplif}) the function $\tilde f(\tilde r,x)$ satisfies a Schr\"odinger like equation with a prescribed asymptotic behavior
\begin{eqnarray}\label{schrod_intro}
\partial_x^2 \tilde f(\tilde r,x) - [x + 2 q^2(x)] \tilde f(\tilde r,x) = - \tilde r \tilde f(\tilde r,x) \;, \; \tilde f(\tilde r,x) \underset{x \to \infty}{\sim} 2^{-1/6} \sqrt{\pi} {\rm Ai}(x-\tilde r) \;.
\end{eqnarray}
We show below that $\tilde f(\tilde r,x)$ is the first component of the $2d-$vector (namely a psi-function) $(\tilde f(\tilde r,x), \tilde g(\tilde r,x))$ which is a solution of the Lax pair associated to the Painlev\'e XXXIV equation \cite{FIKN06} (with parameter $\alpha =0$). Indeed $\tilde f(\tilde r,x)$ and $\tilde g(\tilde r,x)$ are solutions of the system of differential equations
\begin{eqnarray}\label{lax_system_tilde}
\frac{\partial}{\partial \tilde r} 
\left(
\begin{array}{c}
\tilde f(\tilde r,x) \\
\tilde g(\tilde r,x) 
\end{array} \right) 
=  {\mathbf{\tilde A}}\left(
\begin{array}{c}
f(\tilde r,x) \\
g(\tilde r,x) 
\end{array} \right) \;, \;
 \frac{\partial}{\partial x} 
\left(
\begin{array}{c}
\tilde f(\tilde r,x) \\
\tilde g(\tilde r,x) 
\end{array} \right) 
=  {\mathbf{\tilde B}}\left(
\begin{array}{c}
\tilde f(\tilde r,x) \\
\tilde g(\tilde r,x) 
\end{array} \right) \;, \;
\end{eqnarray}
where ${\bf \tilde A}$ and ${\bf \tilde B}$ are $2 \times 2$ matrices given by
\begin{eqnarray}\label{def_AB_tilde}
{\bf \tilde A}
= 
\left( 
\begin{array}{cc}
-\dfrac{q'(x)}{q(x)} & 1+\dfrac{q^2(x)}{\tilde r} \\
-\tilde r-\dfrac{R(x)}{q^2(x)} & \dfrac{q'(x)}{q(x)} 
\end{array}
\right) \;, \;
{\bf \tilde B}
= 
\left( 
\begin{array}{cc}
\dfrac{q'(x)}{q(x)} & -1 \\
\tilde r & - \dfrac{q'(x)}{q(x)} 
\end{array}
\right)\;, \;
\end{eqnarray}
with $R(x)=\int_x^\infty q^2(u)du$ and where the solutions $\tilde f(\tilde r,x)$ and $\tilde g(\tilde r,x)$ are characterized by the asymptotic behaviors \cite{FIKN06}
\begin{subequations}\label{fgtilde_larger_intro}
\begin{eqnarray}
&&\tilde f(\tilde r,x) \underset{\tilde r \to \infty}{\sim} 2^{-1/6} \tilde r^{-1/4} \sin{\left(\frac{2}{3}\tilde r^{3/2} - x \sqrt{\tilde r} + \frac{\pi}{4} \right)} +Ê{\cal O}(\tilde r^{-3/4}) \;, \\
&&\tilde g(\tilde r,x) \underset{\tilde r \to \infty}{\sim} 2^{-1/6} \tilde r^{1/4} \cos{\left(\frac{2}{3}\tilde r^{3/2} - x \sqrt{\tilde r} + \frac{\pi}{4} \right)} +Ê{\cal O}(\tilde r^{-1/4}) \;.
\end{eqnarray}
\end{subequations}
From these expressions, we obtain the asymptotic behaviors of $\tilde \rho_{\rm edge}(\tilde r)$ in Eq. (\ref{rho_intro_simplif}) as
\begin{eqnarray}\label{asympt_tilderho}
\tilde \rho_{\rm edge}(\tilde r) \sim
\begin{cases}
 \frac{1}{2} \tilde r^2 + a_4 \tilde r^4 + {\cal O}(\tilde r^6) \; &, \;  \tilde r \to 0 \\
 \\
 \dfrac{\sqrt{\tilde r}}{\pi} \;&, \; \tilde r \to \infty \;,
\end{cases}
\end{eqnarray}
where $a_4 = -0.196788...$ is given explicitly below (\ref{expr_a4}). Using the large $\tilde r$ behavior of $\tilde \rho_{\rm edge}(\tilde r)$ (\ref{asympt_tilderho}), one can show, as we did before for the density of the eigenvalues $\rho(\lambda,N)$, that the edge and the bulk regimes of $\rho_{\rm DOS}(r,N)$ perfectly match.  

Interestingly, we also show that the mean DOS,  $\rho_{\rm DOS}(r,N)$, is related to the PDF of the gap between the two largest eigenvalues. Let us denote by $\lmax = \Lambda_{1,N} \geq \Lambda_{2,N} \geq \cdots \geq \Lambda_{N,N}$ and by $d_{1,N} = \Lambda_{1,N} - \Lambda_{2,N}$ the first gap. Its PDF is denoted by $p_{\rm GAP}(r,N)$, such that ${{\rm Pr.}} [d_{1,N} \in [r,r+dr]] = p_{\rm GAP}(r,N) dr$. Although the DOS $\rho_{\rm DOS}(N, r)$ in Eq. (\ref{eq:def_rho_eigen}) is naturally defined for $r>0$, our exact formula for finite $N$ suggests a natural analytic continuation to the real negative axis of this function. Furthermore, we show the following identity: 
\begin{eqnarray}\label{id_gap_DOS}
p_{\rm GAP}(r,N) = (N-1) \rho_{\rm DOS}(-r,N) \;.
\end{eqnarray}  
Here we focus on the {\it typical} fluctuations of the gap, i.e. when $r = {\cal O}(N^{-1/6})$ \footnote{{\it Atypically} large fluctuations would correspond to the case where $r = {\cal O}(\sqrt{N})$, which is not studied here.}, and show similarly to Eq. (\ref{DOS_main}) that for large $N$,   
\begin{eqnarray}
p_{\rm GAP}(r,N) = \sqrt{2} N^{1/6} \tilde p_{\rm typ}\left( r\sqrt{2}N^{1/6} \right) \;,
\end{eqnarray}
where $\tilde p_{\rm typ}(\tilde r)$ is given by
\begin{eqnarray}\label{gap_intro_simplif}
\tilde p_{\rm typ}(\tilde r) = \frac{2^{1/3}}{\pi} \int_{-\infty}^\infty \left[\tilde f^2(-\tilde r,x) - \left( \int_x^\infty q(u) \tilde f(-\tilde r,u) du  \right)^2 \right] {\cal F}_2(x) \, dx \;.
\end{eqnarray}
This formula (\ref{gap_intro_simplif}) should be compared to the relatively more complicated formula obtained by WBF in Ref.~\cite{WBF13} (given in Appendix \ref{formula_WBF}). In addition to show a direct connection with the Lax system associated to Painlev\'e XXXIV, our formula (\ref{gap_intro_simplif}) is amenable to a rather precise asymptotic analysis. We show indeed the following asymptotic behaviors:
\begin{eqnarray}\label{asympt_tildegap}
\tilde p_{\rm typ}(\tilde r) = 
\begin{cases}
\frac{1}{2} \tilde r^2 + a_4 \tilde r^4 + {\cal O}(\tilde r^6) \;, \; & r \to 0 \\
\\
A  \; \exp{\left(-\dfrac{4}{3}\tilde r^{3/2} + \dfrac{8}{3}\sqrt{2} \, \tilde r^{3/4}\right)}{\tilde r^{-{21}/{32}}} \left(1 - \dfrac{1405 \sqrt{2}}{1536} \tilde r^{-3/4} + {\cal O}(\tilde r^{-3/2})\right) \;, \; & r \to +\infty \;,
\end{cases}
\end{eqnarray} 
with the amplitude $A = 2^{-91/48} e^{\zeta'(-1)}/\sqrt{\pi}$, where $\zeta'(x)$ is the derivative of the Riemann zeta function, while the amplitude $a_4$ is given in (\ref{expr_a4}). These behaviors should be compared to the distribution of the spacing in the {\it bulk} of the spectrum, given by the Gaudin-Mehta distribution and well approximated by the Wigner surmise \cite{Meh91,For10}, $P_\beta(\tilde r) = A_\beta \tilde r^\beta e^{-B_\beta \tilde r^2}$ (with some constants $A_\beta$ and $B_\beta$). In particular, the latter has a Gaussian tail, at variance with the stretched exponential behavior $\tilde p_{\rm typ}(\tilde r) \sim e^{-\frac{4}{3}\tilde r^{3/2}}$ found here at the {\it edge} (\ref{asympt_tildegap}). 

Note that the occurrence of the Painlev\'e XXXIV equation was previously noted in the study of the OP in Eq. (\ref{def_op_intro}). In particular, using Riemann-Hilbert techniques, it was shown in Ref. \cite{CK08} that the Lax pair studied here enters naturally the expression of the associated Christoffel-Darboux kernel associated to the OP in (\ref{def_op_intro}) in the double scaling limit (see below). More recently, using the Lax method, it was shown in Ref. \cite{AA13} that the Painlev\'e XXXIV hierarchy appears in the present OP system and its generalization to multi-critical matrix models. In this perspective, our main results are thus to show that (i) near extreme statistics, including order statistics, of GUE can naturally be expressed in terms of the OP given in Eq. (\ref{def_op_intro}) and (ii) provide a rather simple study of this OP system (\ref{def_op_intro}) in the double scaling limit, along the lines developed in theoretical physics \cite{NM11,PS90,GM94,Sch12}, where we will see that the Lax pair associated to Painlev\'e XXXIV emerges rather naturally. 

The paper is organized as follows: in section 2, we give an explicit formula for $\rho_{\rm DOS}(r,N)$ for any finite $N$ in terms of the OP system in (\ref{def_op_intro}). In section 3 we perform a large $N$ analysis of $\rho_{\rm DOS}(N,r)$, both in the bulk and in the edge scaling limit -- the latter corresponding to a double scaling limit of the OP system. In section 4, we focus on the typical fluctuations of the first gap, where $p_{\rm GAP}(r,N)$ is analyzed in the limit when $N$ is large and $r \sim {\cal O}(N^{-1/6})$. In section 5, we present a numerical evaluation of our formulas, providing in particular a direct comparison with the result of WBF \cite{WBF13}, before we conclude in section 5. Some technical details have been left in Appendices A, B and C while the formula of WBF \cite{WBF13} for the PDF of the first gap has been given, for completeness, in Appendix~D.

\section{An exact formula for $\rho_{\rm DOS}(r,N)$ for finite $N$}

We start by deriving an exact formula for the density of states $\rho_{\rm DOS}(r,N)$
valid for any finite $N$, in terms of the OP in (\ref{def_op_intro}). The obtained formula turns out to be useful for a large $N$ asymptotic analysis. 
\subsection{Introducing orthogonal polynomials}

\begin{figure}
\begin{center}
\psfrag{A}[l][][3]{$p_{\rm GAP}(r,4)$}
\psfrag{B}[l][][3]{$3\,\rho_{\rm DOS}(r,4)$}
\psfrag{r}[c][][3]{$r$}
\psfrag{0}[c][][2.5]{$0$}
\psfrag{1}[c][][2.5]{$1$}
\psfrag{5}[c][][2.5]{$5$}
\rotatebox{-90}{\resizebox{65mm}{!}{\includegraphics{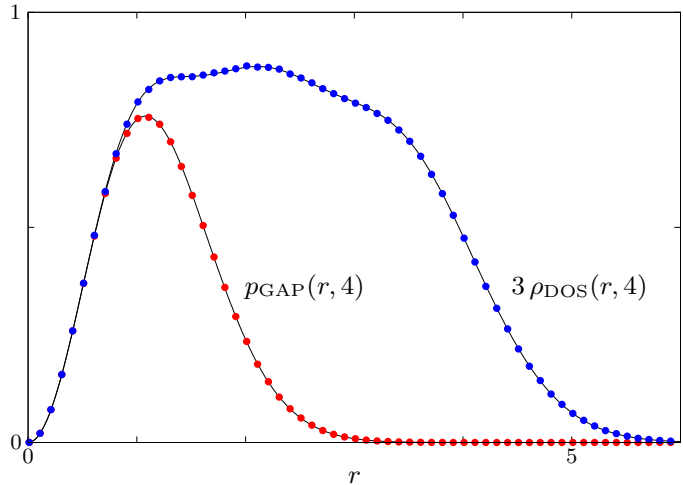}}}
\caption{Numerical result for a $4\times4$ GUE matrix (dots) compared to our analytical results -- see Eqs. (\ref{id_gap_DOS}) and (\ref{rho_kernel}) -- (black curve).}
\label{example_N_4}
\end{center}
\end{figure}

First, we notice that, using the invariance of the joint PDF $P_{\rm joint}(\lambda_1, \cdots, \lambda_N)$ in (\ref{jPDF}) under any permutation of the $\lambda_i$'s, the mean DOS $\rho_{\rm DOS}(r,N)$ can be written as
\begin{equation}\label{rho_start}
\rho_{\rm DOS}(r,N)=N \int_{-\infty}^{\infty} {\rm d}y
\int_{-\infty}^{y} {\rm d}\lambda_{1} \int_{-\infty}^{y} {\rm d}\lambda_2\, \cdots \int_{-\infty}^{y} {\rm d}\lambda_{N-2}\,
P_{\rm joint}(\lambda_1,\lambda_2,...,\lambda_{N-2},y-r,y) \;,
\end{equation}
where $y$ denotes the actual value of $\lmax$. To perform this multiple integral in (\ref{rho_start}) it is convenient to introduce the monic orthogonal polynomials which were briefly presented in the introduction (\ref{def_op_intro}) $\pi_k(\lambda,y)$ (which are polynomials of degree $k$ of the variable $\lambda$, while $y$ is a parameter) defined by \cite{NM11}
\begin{eqnarray}\label{def_op}
\left\{
     \begin{array}{lr}
			\langle \pi_k, \pi_{k'}\rangle =  \int_{-\infty}^{y} {\rm d}\lambda \, \pi_k(\lambda,y) \pi_{k'}(\lambda,y) e^{-\lambda^2} = \delta_{k,k'}h_k(y)\\
			\\
			\pi_k(\lambda,y)=\lambda^k+...
     \end{array}
\right.
\end{eqnarray}
When $y \to \infty$, these orthogonal polynomials reduce to the Hermite polynomials. For finite $y$, there is no closed formula for these orthogonal polynomials. The first polynomials $\pi_k(\lambda,y)$ can however be computed from (\ref{def_op}), to obtain \cite{NM11}:
\begin{eqnarray}\label{first_op}
\pi_0(\lambda,y)&=&1\\
\pi_1(\lambda,y)&=&\lambda+a\nonumber\\
\pi_2(\lambda,y)&=&\lambda^2 + \lambda \left(\frac{a + \lambda}{1 - 2 a (a + \lambda)} - \lambda\right) - 1 + \frac{1}{2-4 a (a + \lambda)}\nonumber\\
\pi_3(\lambda,y)&=&\lambda^3
+\left[\lambda^2 2 a \left(8 a^2 - 4 a^2 y^2+ 8 a y - 8 a y^3 - 3- 4 y^4\right) \right. \nonumber \\
&&+2 \lambda \left(-12 a^3 y-10 a^2- 22 a^2 y^2 -9ay- 10 a y^3+3\right)\nonumber\\
&&\hspace*{-1.5cm}+\left.a \left(- 16 a^2 + 4 a^2 y^2- 20 a y + 8 a y^3 +5- 4 y^2+ 4 y^4\right)\right]/\left(8 a^3 y+12 a^2 + 16 a^2 y^2 + 12 a y + 8 a y^3-4\right)\nonumber
\end{eqnarray}
where $a \equiv a(y)$ is given by
\begin{eqnarray}
a = \frac{e^{-y^2}}{\sqrt{\pi} (1 + \text{erf}(y))} \;,
\end{eqnarray} 
where ${\rm erf}(y)$ denotes the error function. From (\ref{first_op}) one also obtains the first norms
\begin{eqnarray}\label{first_ampli}
h_0&=& e^{-y^2}\frac{1}{2a}\\
h_1&=&e^{-y^2} \frac{1}{4}  (\frac{1}{a}-2 a-2 y)\nonumber\\
h_2&=&e^{-y^2} \frac{2 a^3 y+3 a^2 + 4 a^2 y^2+ 3 a y + 2 a y^3 - 1}{
4 a (2 a^2 + 2 a y - 1)}\nonumber\\
h_3&=&e^{-y^2} \frac{- 32 a^4+ 4 a^4 y^2 - 60 a^3 y+ 16 a^3 y^3+29 a^2  - 20 a^2 y^2+ 20 a^2 y^4 + 30 a y  + 8 a y^3  + 8 a y^5 -6}{16 a (2 a^3 y +3 a^2+ 4 a^2 y^2 + 3 a y + 2 a y^3 - 1)}\nonumber
\end{eqnarray}
As one sees in Eq. (\ref{first_op}), the expression of $\pi_k(\lambda,y)$ becomes more and more complicated as $k$ grows and to analyze these polynomials for larger values of $k$, it is much more convenient to consider the three term recurrence relation which they satisfy \cite{szego}
\begin{eqnarray}
&&\lambda \, \pi_k(\lambda,y) = \pi_{k+1}(\lambda,y) + S_k(y) \pi_k(\lambda,y) + R_k(y) \pi_{k-1}(\lambda,y) \;, \label{recurrence_op}\\
&&{\rm with} \; R_k(y) = \frac{h_k(y)}{h_{k-1}(y)}  \label{def_Rk} \;,
\end{eqnarray} 
and $S_k(y) \neq 0$ as the interval of integration in (\ref{def_op}) is not symmetric. The density $\rho_{\DOS}(r,N)$ in (\ref{rho_start}) can then be expressed in terms of these orthogonal polynomials $\pi_k(\lambda,y)$. This is done by (i) replacing the Vandermonde determinant in $P_{\rm joint}$ (\ref{jPDF}, \ref{rho_start}) by a determinant built from the OP $\pi_k(\lambda,y)$ and (ii) writing each determinant in terms of its Laplace expansion and use the orthogonality condition (\ref{def_op}) to perform the integrals over $\lambda_{1}, \cdots, \lambda_{N-2}$ in (\ref{rho_start}). By performing these standard manipulations \cite{Meh91,For10}, we find that $\rho_{\DOS}(r,N)$ can be written as
\begin{eqnarray}\label{rho_kernel}
\rho_{\DOS}(r,N)=\frac{N(N-2)!}{Z_N} \int_{-\infty}^{\infty} {\rm d}y\,
\prod_{k=0}^{N-1}h_k(y)
\left|
\begin{array}{ccc}
K_N(y-r,y-r) & K_N(y-r,y) \\
K_N(y,y-r) & K_N(y,y) 
\end{array}
\right| \;,
\end{eqnarray}
where $K_N(\lambda,\lambda')$ is the kernel associated to the OP in (\ref{def_op}), given by
\begin{eqnarray}
&&K_N(\lambda,\lambda')=\sum_{k=0}^{N-1} \frac{1}{h_k(y)}\pi_k(\lambda,y)\pi_k(\lambda',y)  e^{-\frac{\lambda^2+\lambda'^2}{2}} = \sum_{k=0}^{N-1} \psi_k(\lambda,y)\psi_k(\lambda',y) \;, \label{kernel_1} \\
&&{\rm where \;} \; \psi_k(\lambda,y) = \frac{1}{\sqrt{h_k(y)}}\pi_k(\lambda,y) e^{-\frac{\lambda^2}{2}} \;, \label{def_psi}
\end{eqnarray}
where we have introduced the normalized wave functions $\psi_k(\lambda,y)$. Using the Cristoffel-Darboux formula, the kernel $K_N(\lambda,\lambda')$ in (\ref{kernel_1}) can be written as
\begin{eqnarray}
K_N(\lambda,\lambda') = \sqrt{R_N(y)} \frac{\psi_N(\lambda,y) \psi_{N-1}(\lambda',y) - \psi_{N-1}(\lambda,y) \psi_{N}(\lambda',y)}{\lambda-\lambda'} \;, \label{kernel_CD} 
\end{eqnarray}
which, of course, also depends on $y$. We note in passing that, although $\rho_{\rm DOS}(r,N)$ is naturally defined for $r>0$ [see Eq. (\ref{def_rho})], the above expression (\ref{rho_kernel}) admits a natural extension for $r<0$. This formula~(\ref{rho_kernel}) can be explicitly evaluated for small values of $N$ using the expressions in Eqs. (\ref{first_op}) and (\ref{first_ampli}). For instance, for $N=4$, using these explicit expressions 
one obtains the expression of $\rho_{\rm DOS}(r, 4)$ as a single integral which 
%\begin{eqnarray}\label{rho_4}
%&&\rho_{\rm DOS}(r,4)=\int_{-\infty}^{\infty}{\rm d}y\frac{e^{-4 y^2+2 y r-r^2} }{72 a^2 \pi ^2}r^2 (45+4 (-32 a^2-33 a y+21 a^2 y^2+50 a y^3+30 y^4\nonumber\\&&+12 a^2 y^4+28 a y^5+16 y^6+4 a^2 y^6+8 a y^7+4 y^8-2 (30 y^3+24 y^5+8 y^7\nonumber\\&&+2 a^2 y (13+8 y^2+4 y^4)+a (3+54 y^2+40 y^4+16 y^6)) r\nonumber\\&&+2 (-3+3 y^2 (7+8 y^2+4 y^4)+2 a^2 (8+5 y^2+6 y^4)+a y (33+34 y^2+24 y^4)) r^2\nonumber\\&&-4 (a+(3-2 a^2) y+2 a y^2+4 (1+a^2) y^3+8 a y^4+4 y^5) r^3\nonumber\\&&+(3+4 (a+y) (y^3+a (-2+y^2))) r^4)) \;.
%\end{eqnarray}
%The remaining integral in (\ref{rho_4}) 
can easily be computed numerically. Eventually, as a check of our computation, this exact analytical expression can be compared to a numerical evaluation of this quantity, obtained by sampling a large number of $4 \times 4$ matrices. The result of this comparison is shown in Fig. \ref{example_N_4} which shows a very nice agreement between our exact formula (\ref{first_op}, \ref{first_ampli}, \ref{rho_kernel}) -- the black line on the figure -- and the numerics -- the blue points on that figure. We also notice that, compared to the density of eigenvalues $\rho(\lambda,N)$ (\ref{def_rho}), the DOS $\rho_{\rm DOS}(r,N)$, for finite $N$, shows much less pronounced oscillations.    
This is also quite visible on Fig. \ref{fig_illust_DOS} for larger values of $N$. 

To conclude this paragraph, we emphasize that the exact expressions in (\ref{rho_kernel}) together with (\ref{kernel_CD}), which are valid for any value of $N$, constitute the starting point of the asymptotic large $N$ analysis which we will perform in section 3. Before that, we derive a useful identity.

\subsection{A useful identity and the normalization of $\rho_{\DOS}(r,N)$}

At this stage, it is useful to notice the following identity, whose derivation starts with the definition of the cumulative distribution of the largest eigenvalue $F_N(y) = {\rm Pr.}(\max_{1 \leq i \leq N} \lambda_i \leq y)$: 
\begin{eqnarray}
F_N(y) = \int_{-\infty}^{y} d\lambda_1 \cdots  \int_{-\infty}^{y} d\lambda_N P_{\rm joint}(\lambda_1, \cdots, \lambda_N) \;.
\end{eqnarray}
It can be evaluated by replacing the Vandermonde determinant by the determinant built from the OP in (\ref{def_op}) and then use the Cauchy-Binet formula to obtain \cite{Meh91,For10,NM11}
\begin{eqnarray}\label{identity_1}
F_N(y) = \frac{N!}{Z_N} \prod_{j=0}^{N-1} h_j(y) \;. 
\end{eqnarray}
 Differentiating Eq. (\ref{identity_1}) with respect to $y$, one obtains
\begin{equation}\label{identity_2}
{\partial_y} F_N(y)  =  \frac{N!}{Z_N} \sum_{k=0}^{N-1} \left( {\partial_y h_k(y)} \prod_{j= 0, j \neq k}^{N-1} h_j \right)= \frac{N!}{Z_N}  \left(\prod_{j= 0}^{N-1} h_{j}(y)\right)\sum_{k=0}^{N-1} \frac{\partial_y h_k(y)}{h_k(y)}  \;.
\end{equation}
On the other hand, one has from (\ref{def_op})
\begin{eqnarray}
{\partial_y h_k(y)} = [\pi_k(y,y)]^2 e^{-y^2} + 2 \langle \partial_{y} \pi_k, \pi_k\rangle =  [\pi_k(y,y)]^2 e^{-y^2} \;,
\end{eqnarray}
as $\partial_{y} \pi_{k}(\lambda,y)$ is a polynomial in $\lambda$ of degree $k-1$ [because $\pi_k(\lambda,y)$ is a monic polynomial of the variable $\lambda$ (\ref{def_op})], which implies $\langle \partial_{y} \pi_k, \pi_k\rangle=0$. Hence
\begin{eqnarray}\label{identity_3}
\partial_{y} \ln{F_N(y)} = \sum_{k=0}^{N-1} \frac{[\pi_k(y, y)]^2}{h_k(y)}e^{-y^2} = K_N(y,y) \;.
\end{eqnarray}
Using (\ref{identity_1}) and (\ref{identity_3}), one can write $\rho_{\DOS}(r,N)$ in (\ref{rho_kernel}) as
\begin{eqnarray}\label{rho_kernel_2}
\rho_{\rm DOS}(r,N) = \frac{1}{N-1} \int_{-\infty}^\infty d y \left[ F'_N(y) K_N(y-r,y-r) - F_N(y) K^2_N(y,y-r)  \right],
\end{eqnarray}
where we have used that the kernel is symmetric $K_N(\lambda,\lambda') = K_N(\lambda',\lambda)$. Note that the second term in (\ref{rho_kernel_2}) is a contribution coming from the correlations between the largest eigenvalue $\lmax$ in $y$ and the eigenvalues at $y - r$. One can finally check the normalization of $\rho_{\rm DOS}(r,N)$ by using the identities
\begin{eqnarray}
&&\int_0^{+\infty} K_N(y-r, y-r) dr = \int_{-\infty}^{y} K_N(x,x) dx = N \;, \label{normalization1} \\
&& \int_0^{+\infty}K^2_N(y,y-r) dr = K_N(y,y) = \partial_{y} \ln{F_N(y)} \label{normalization2} \;,
\end{eqnarray}
where we have used the identity (\ref{identity_3}) in (\ref{normalization2}). Note that from (\ref{normalization1}, \ref{normalization2}) we see that 
the weight of the first term in (\ref{rho_kernel_2}) is of order ${\cal O}(1)$ (for large $N$) while the second term in (\ref{rho_kernel_2}) is of order ${\cal O}(N^{-1})$. We now proceed to the large $N$ analysis of $\rho_{\DOS}(r,N)$.

\section{Large $N$ asymptotics}

As we have already pointed it out in the introduction [see for instance Eq. (\ref{scaling_density})], there are two different regimes characterizing the fluctuations of the eigenvalues:
one is the {\it bulk} regime, for eigenvalues $\lambda_i$'s such that $\lambda_i/\sqrt{N} = {\cal O}(1)$ and $\lambda_i < \sqrt{2N}$, and the other scaling regime is at the {\it edge} where the eigenvalues are close to the edge of the Wigner semi-circle, $\lambda - \sqrt{2N} = {\cal O}(N^{-1/6})$. Hence when analyzing the large $N$ behavior of $\rho_{\DOS}(r,N)$, one expects two different scaling regimes: (i) when $r = {\cal O}(\sqrt{N})$ and (ii) when $r = {\cal O}(N^{-1/6})$. We now study these two regimes separately.

\subsection{Regime (i): in the bulk when $r = {\cal O}(\sqrt{N})$}

In this regime, it is easy to see that the second term in (\ref{rho_kernel_2}) is subdominant compared to the first one. We have already seen before that the total weight of the latter is only of order ${\cal O}(N^{-1})$ while the first term has a total weight of order ${\cal O}(1)$ [see our comment below Eqs. (\ref{normalization1}, \ref{normalization2})]. Physically this is also clear as the second one involves the correlations between $\lmax$, located in $y$, and the eigenvalues located in 
$y - r$, where in this regime (i), $\lmax - r = {\cal O}(\sqrt{N})$. To analyze the first term in (\ref{rho_kernel_2}) one can replace the PDF of $\lmax$, ${\rm Pr.}(\lmax = y) = F'_N(y)$ by $\delta(y- \sqrt{2N})$:  this is justified by the fact that $K_N(y - r,y-r)$ with $y-r = {\cal O}(\sqrt{N})$ is not sensitive to the typical fluctuations of $\lmax$ around $y = \sqrt{2N}$ which are much smaller, of order ${\cal O}(N^{-1/6})$. Denoting $r = \hat r \sqrt{N}$, one thus
has
\begin{eqnarray}\label{shifted_wigner}
&&\rho_{\rm DOS}(\hat r \sqrt{N},N) \sim \frac{1}{N} K_N(\sqrt{2N} - \hat r \sqrt{N}, \sqrt{2N} - \hat r \sqrt{N}) \underset{N \to \infty}{\sim} \frac{1}{\sqrt{N}} \tilde \rho_{\rm bulk}\left(\hat r = \frac{r}{\sqrt{N}}\right) \;,\nonumber \\
&&\tilde \rho_{\rm bulk}(\hat r) = \rho_W(\sqrt{2}-\hat r)=\frac{1}{\pi} \sqrt{\hat r(2\sqrt{2} - \hat r)} \;,
\end{eqnarray}
which yields the result announced in the first line of (\ref{DOS_main}) and in Eq. (\ref{eq:tilde_bulk}). As explained in the introduction, this result (\ref{shifted_wigner}) can be easily understood as the density of eigenvalues at a distance $r$ from $\lmax \sim \sqrt{2N}$ in (\ref{eq:def_rho_eigen}) such that $r - \lmax = {\cal O}(\sqrt{N})$ is insensitive to the fluctuations of $\lmax$ which are of order ${\cal O}(N^{-1/6})$. Hence, $\rho_{\DOS}(r,N)$ is simply a shifted Wigner semi-circle (\ref{shifted_wigner}). Note that this argument holds actually for any Gaussian $\beta$-ensemble where the joint PDF of the eigenvalues $\lambda_i$'s is given by Eq. (\ref{jPDF}) with the replacement of $\prod_{i,j}|\lambda_i-\lambda_j|^2$ by $\prod_{i,j}|\lambda_i-\lambda_j|^\beta$, where $\beta > 0$ can take any real value \cite{Meh91,For10}. In this case, one thus expects that the DOS behaves also as in Eq. (\ref{shifted_wigner}). We now turn to the case where $r$ is of order ${\cal O}(N^{-1/6})$.

\subsection{Regime (ii): near the edge when $r = {\cal O}(N^{-1/6})$}

In this regime, the analysis is more subtle. It requires the analysis of the kernel $K_N(y - r,y-r')$ when $y \sim \sqrt{2N}$ and both $r,r'$ are of order $N^{-1/6}$. This corresponds to a double scaling limit analysis of the recursion relations for the OP in Eqs. (\ref{recurrence_op}) and (\ref{def_Rk}). In Ref. \cite{NM11}, Nadal and Majumdar used this OP system (\ref{def_op}) to compute the cumulative distribution function (CDF) of $\lmax$ in Eq. (\ref{identity_1}), which can be expressed in terms of the norms $h_k(y)$'s only. To derive the Tracy-Widom distribution from (\ref{identity_1}), one needs indeed to perform the analysis of $h_k(y)$ for large $k$ of order ${\cal O}(N)$, with $N \to \infty$ and, simultaneously,  $y - \sqrt{2N} = {\cal O}(N^{-1/6})$. This double scaling analysis of the norms $h_k(y)$ was performed in Ref. \cite{NM11} and generalized to the multi critical matrix models in Ref. \cite{AA13}. Here, in addition to the norms, the computation of $\rho_{\DOS}(r)$ in (\ref{rho_kernel}) requires the analysis of the kernel $K_N(y-x,y-x')$ in this double scaling limit when $N \to \infty$, with $y - \sqrt{2N} = {\cal O}(N^{-1/6})$ and with $x$ and $x'$ both of order ${\cal O}(N^{-1/6})$. To this purpose, we analyze and solve the three term recurrence relation in (\ref{recurrence_op}) in this double scaling limit.

\subsubsection{Double scaling analysis of the three terms recurrence relation}

We start with the recursion relation satisfied by the wave functions $\psi_N(\lambda,y)$ in Eq. (\ref{def_psi}), which is easily obtained from the recursion relation for the OP given in Eq. (\ref{recurrence_op}). It reads 
\begin{eqnarray}\label{recurrence_psi}
\lambda \, \psi_N(\lambda,y) = \sqrt{R_{N+1}(y)} \psi_{N+1}(\lambda,y) +S_N(y) \psi_N(\lambda,y)+\sqrt{R_N(y)}\psi_{N-1}(\lambda,y) \;,
\end{eqnarray}
where we recall that $\psi_N(\lambda,y)={\pi_N(\lambda,y)}/{\sqrt{h_N}}e^{-\frac{\lambda^2}{2}}$, $R_N={h_N}/{h_{N-1}}$ and $S_N=-{\partial_y h_N(y)}/2$.
From Ref. \cite{NM11}, we know that, in the double scaling limit when $N \to \infty$ and $y -\sqrt{2N} = {\cal O}(N^{-1/6})$ the coefficients $R_N(y)$ and $S_N(y)$ take the scaling form
\begin{eqnarray}\label{double_scaling_NM}
R_N(y)&=&\frac{N}{2}\left(1-N^{-\frac23}\, q^2(x)+{\cal O}(N^{-1}) \right) \;, \; S_N(y)=-\frac{N^{-\frac16}}{\sqrt{2}} q^2(x)+{\cal O}(N^{-\frac12}) \;, \\
y&=&\sqrt{2N}+\frac{x}{\sqrt{2}}N^{-\frac16} \;, \nonumber 
\end{eqnarray}
where $q(x)$ is the Hastings-McLeod solution of the Painlev\'e II equation (\ref{hasting}). Given the scaling form for $R_N(y)$ and $S_N(y)$ in (\ref{double_scaling_NM}), we search for a solution of the recursion relations in (\ref{recurrence_psi}) under the scaling form 
\begin{eqnarray}\label{scaling_psi}
\psi_N(y-r,y) = \frac{2^{1/4}}{\sqrt{\pi}}N^{\nu} G\left(\sqrt{2} N^{1/6} r, \sqrt{2}N^{1/6}(y - \sqrt{2N}) \right) \;,
\end{eqnarray}
where $\nu$ and the function $G$ are still to be determined and where the amplitude $2^{1/4}/\sqrt{\pi}$ is chosen here for convenience. As mentioned above, the choice of the dependence of $\psi_N(y-r,y)$ in the scaling variable $\sqrt{2}N^{1/6}(y - \sqrt{2N})$ follows naturally from the dependence of the $R_N(y)$ and $S_N(y)$ in (\ref{recurrence_psi}), (\ref{double_scaling_NM}) \cite{NM11}. The dependence on the scaling variable $\sqrt{2} N^{1/6} r$ is a priori less obvious on the recurrence relation itself (\ref{recurrence_psi}). It is however motivated by the analysis of the OP system (\ref{def_op}) in the limit where $y \gg \sqrt{2N}$. Indeed, for $y \gg \sqrt{2N}$, one can replace $y$ by $+ \infty$ in the integral defining the OP $\pi_k(\lambda,y)$ in (\ref{def_op}) such that the OP can be expressed in terms of the Hermite polynomials. This yields, for $y \gg \sqrt{2N}$ \cite{NM11}:
\begin{eqnarray}\label{psi_hermite}
\psi_N(\lambda,y) = \frac{1}{\pi^{1/4} 2^{N/2} \sqrt{N!}} H_N(\lambda) e^{-\frac{\lambda^2}{2}} + {\cal O}(e^{-y^2})\;, 
\end{eqnarray}
where $H_N(\lambda)$ is the Hermite polynomial of degree $N$. We now want to evaluate the wave function $\psi_N(y-r,y)$ when $y$ reaches the value $\sqrt{2N}$ from above, $y \to \sqrt{2N}^+$. Setting $y = \sqrt{2N} + x N^{-1/6}/\sqrt{2}$ as in (\ref{double_scaling_NM}) one finds, from (\ref{psi_hermite}), using the Plancherel-Rotach formula for Hermite polynomials \cite{szego} that $\psi_N(y-r,y)$ approaches the limiting form 
\begin{eqnarray}\label{Plancherel_Rotach}
\psi_N(y  -r,y) \underset{N \to \infty}\sim 2^{1/4} N^{-1/12} {\rm Ai}(x-\sqrt{2}N^{1/6}r) +  {\cal O}(e^{-y^2}) \;,
\end{eqnarray}
which is thus a function of the two scaling variables $x$ in (\ref{double_scaling_NM}) and 
\begin{eqnarray}\label{r_tilde}
\tilde r = \sqrt{2}N^{1/6}r \;,
\end{eqnarray}
as proposed in (\ref{scaling_psi}). Furthermore,  
by matching the two formulas (\ref{scaling_psi}) and (\ref{Plancherel_Rotach}) one finds
\begin{eqnarray}\label{matching}
\nu = -\frac{1}{12} \;, \; G(\tilde r,x) \underset{x \to \infty}{\sim} \sqrt{\pi}{\rm Ai}(x-\tilde r) \;.
\end{eqnarray}
The next step is then to insert the ansatz (\ref{scaling_psi}) into the recursion relation (\ref{recurrence_psi}) and then perform a large $N$ expansion
in the double scaling limit, corresponding to the regime in (\ref{double_scaling_NM}). For this purpose we will also need the large $N$ expansion of the quantities
$x_N = \sqrt{2}N^{1/6}(y- \sqrt{N})$ and $r_N = \sqrt{2}N^{1/6}r$:
\begin{eqnarray}\label{xN_rN}
x_{N+1} = x_N - N^{-1/3} + {\cal O}(N^{-1}) \;, r_{N+1} = r_N  + {\cal O}(N^{-1}) \;.
\end{eqnarray}
After some straightforward algebra, one finds (remembering that $r = \tilde r N^{-1/6}/\sqrt{2}$) the limiting form of the kernel in the double scaling limit, from (\ref{kernel_CD}) and (\ref{scaling_psi}):
\begin{eqnarray}\label{kernel_dble_scaling}
K_N(y - r, y -r') \underset{N \to \infty}{\sim} N^{1/6}\sqrt{2} \frac{G(\tilde r,x) \partial_x G(\tilde r',x)- \partial_x G(\tilde r,x) G(\tilde r',x)}{\pi(\tilde r' - \tilde r)} \;,
\end{eqnarray}
where $G(\tilde r,x)$ is the solution of the Schr\"odinger equation:
\begin{eqnarray}\label{eq:schrod}
-\partial^2_x G(\tilde r,x) + [x + 2q^2(x)] G(\tilde r,x) =  \tilde r \, G(\tilde r,x) \;,  
\end{eqnarray}
with the asymptotic behavior given in Eq. (\ref{matching}) and where $q(x)$ in (\ref{eq:schrod}) is the Hastings-McLeod solution of Painlev\'e II (\ref{hasting}).

In view of future purpose, it is also useful to study the asymptotic behavior of  $G(\tilde r,x)$ when both $\tilde r$ and $x$ are large. If one assumes that the large $\tilde r$ limit and the large $x$ limit do commute, this asymptotic behavior can be obtained from Eq. (\ref{matching}) where one takes the large $\tilde r$ limit. Using the asymptotic behavior of the Airy function for large negative argument, one obtains
\begin{eqnarray}\label{g_largerx}
G(\tilde r,x) {\sim}  \frac{1}{\tilde r^{1/4}} \sin{\left( \frac{2}{3}\tilde r^{3/2} - x \sqrt{\tilde r} +\frac{\pi}{4}  \right)}  \;, \; {\rm for} \; \tilde r \gg 1 \; \& \; x \gg 1 \;,
\end{eqnarray}
which will be useful in the following to solve the equation for $G(\tilde r,x)$ which is the purpose of the next section.

\subsubsection{Solution of the Schr\"odinger equation and the Lax pair of Painlev\'e XXXIV}

We first notice that, if we set $\tilde r=0$ in (\ref{eq:schrod}), we see, given the large $x$ behavior (\ref{matching}) that $G(0,x)$ coincides, up to a constant, with $q(x)$ in Eq. (\ref{hasting})
\begin{eqnarray}\label{G_r0}
G(\tilde r = 0,x) = \sqrt{\pi} q(x) \;,
\end{eqnarray}
which is already an interesting result. The second observation is that the Schr\"odinger equation (\ref{eq:schrod}) has a supersymmetric structure. This property follows from the fact that $q(x)$ is solution of the Painlev\'e II equation (\ref{hasting}), which implies
\begin{eqnarray}\label{qqprime}
&&x + 2 q^2(x) = \frac{q''(x)}{q(x)} = \frac{d}{dx} \left(\frac{q'(x)}{q(x)} \right) + \left(\frac{q'(x)}{q(x)} \right)^2 = -Q'(x) + Q^2(x) \;, \\
&& {\rm with \;} Q(x) = - \frac{q'(x)}{q(x)} \;.
\end{eqnarray}
The next step is to realize that $Q(x) = - {q'(x)}/{q(x)}$ can actually be expressed in terms of the Hastings-McLeod solution of the Painlev\'e II equation with parameter $\alpha = 1/2$, namely (see for instance Ref.~\cite{WBF13,CJP99})
\begin{eqnarray}\label{relQ_q1/2}
Q(x) =  2^{1/3} q_{1/2}(-2^{1/3} x) \;,
\end{eqnarray}
where $q_{1/2}(s)$ satisfies
\begin{eqnarray}\label{P2_alpha1/2}
q_{1/2}''(s) = 2q_{1/2}^3(s) + s q_{1/2}(s) - \frac{1}{2} \;,
\end{eqnarray}
with the asymptotic behavior (characterizing the Hastings-McLeod solution in this case):
\begin{eqnarray}\label{asympt_alpha1/2}
q_{1/2}(s) \underset{s\to \infty}{\sim} \frac{1}{2s} \;,
\end{eqnarray}
while for large negative argument one has
\begin{eqnarray}\label{asympt_alpha1/2_neg}
q_{1/2}(s) \underset{s\to -\infty}{\sim} \sqrt{-\frac{s}{2}} \;.
\end{eqnarray}
From the definition of $Q(x) = -q'(x)/q(x)$ together with the definition of $q(x)$ (\ref{hasting}) it is indeed straightforward to check that $q_{1/2}(s = -2^{-1/3}x)$ defined through Eq. (\ref{relQ_q1/2}) satisfies the Painlev\'e II equation (\ref{P2_alpha1/2}) together with the asymptotic behavior in (\ref{asympt_alpha1/2}). Hence, using Eqs. (\ref{qqprime}), (\ref{relQ_q1/2}) one can express $G(\tilde r,x)$ as
\begin{eqnarray}\label{schrod_alpha1/2}
G(\tilde r,x) = \phi(\tilde r,-2^{1/3} x) \;, \; {\rm where \;} \partial^2_s \phi(\tilde r,s)  - \left[q_{1/2}'(s) + q_{1/2}^2(s)\right] \phi(\tilde r,s) = - \frac{\tilde r}{2^{2/3}} \phi(\tilde r,s) \;,
\end{eqnarray}
with the appropriate asymptotic behavior of $\phi(\tilde r,s)$ deduced from (\ref{matching}):
\begin{eqnarray}\label{asympt_phi}
\phi(\tilde r,s) \underset{s \to - \infty}{\sim} \sqrt{\pi} {\rm Ai}(-2^{-1/3} s - \tilde r) \;.
\end{eqnarray}

The final step is to relate $\phi(\tilde r,s)$, which satisfies this supersymmetric Schr\"odinger equation~(\ref{schrod_alpha1/2}), to a solution of the Lax pair associated to a particular Painlev\'e XXXIV equation~\cite{CK08}. A similar strategy was used in Ref. \cite{Sch12} to solve a related supersymmetric Schr\"odinger equation. The relevant Lax pair for our problem is the following one \cite{CK08,FIKN06}
\begin{eqnarray}\label{lax_system}
\frac{\partial}{\partial r} 
\left(
\begin{array}{c}
f(r,s) \\
g(r,s) 
\end{array} \right) 
=  {\mathbf A}\left(
\begin{array}{c}
f(r,s) \\
g(r,s) 
\end{array} \right) \;, \;
 \frac{\partial}{\partial s} 
\left(
\begin{array}{c}
f(r,s) \\
g(r,s) 
\end{array} \right) 
=  {\mathbf B}\left(
\begin{array}{c}
f(r,s) \\
g(r,s) 
\end{array} \right) \;, \;
\end{eqnarray}
where ${\bf A}$ and ${\bf B}$ are $2 \times 2$ matrices given by
\begin{eqnarray}\label{def_AB}
{\bf A}
= 
\left( 
\begin{array}{cc}
2 \, q_{1/2}(s) & 2 + \left[q_{{1}/{2}}^2(s) + q_{1/2}'(s) + \tfrac{s}{2}\right]/r \\
- 2 r - q_{1/2}^2(s)+q_{1/2}'(s) - \tfrac{s}{2} & -2 \, q_{1/2}(s) 
\end{array}
\right) \,, \,
{\bf B}
= 
\left( 
\begin{array}{cc}
q_{1/2}(s) & 1 \\
- r & -q_{1/2}(s) 
\end{array}
\right), 
\end{eqnarray}
where the solutions $f(r,s)$ and $g(r,s)$ are characterized by the asymptotic behavior {\it on the positive real axis} (note in particular that for large negative $\tilde r$, the asymptotic behaviors are different \cite{CK08}, see also below)~\cite{CK08,FIKN06}
\begin{subequations}\label{fg_larger}
\begin{eqnarray}
&&f(r,s) = r^{-1/4} \sin{\left(\frac{4}{3}r^{3/2} + s r^{1/2} + \frac{\pi}{4} \right)} + {\cal O}(r^{-3/4})  \\
&&g(r,s) = r^{1/4} \cos{\left(\frac{4}{3}r^{3/2} + s r^{1/2} + \frac{\pi}{4} \right)} + {\cal O}(r^{-1/4}) \;.
\end{eqnarray}
\end{subequations}

The connection between $\phi(\tilde r,s)$ and the solution of this Lax system in Eqs. (\ref{lax_system}, \ref{def_AB}) is through the function $f(r,s)$. It is indeed easy to show from Eqs. (\ref{lax_system}, \ref{def_AB}) that $f(r,s)$ satisfies the following Schr\"odinger equation
\begin{eqnarray}\label{eq_f}
\partial^2_s f(r,s)  - \left[q_{1/2}'(s) + q_{1/2}^2(s)\right] f(r,s) = - r f(r,s) \;,
\end{eqnarray} 
which is exactly similar to the equation satisfied by $\phi(\tilde r,s)$ in Eq. (\ref{schrod_alpha1/2}) with the substitution $r = 2^{-2/3}\tilde r$. One can study the large negative $s$ behavior of $f(r,s)$ by plugging the large negative $s$ behavior of $q_{1/2}(s)$ (\ref{asympt_alpha1/2_neg}) in Eq. (\ref{eq_f}). This yields:
\begin{eqnarray}
\partial^2_s f(2^{-2/3}\tilde r,s) + \frac{s}{2}f(2^{-2/3}\tilde r,s) = -2^{-2/3}\tilde r f(2^{-2/3}\tilde r,s) \;, \; {\rm when} \; s \to - \infty \;.
\end{eqnarray} 
Hence, one gets that for $s \to -\infty$, 
\begin{eqnarray}
f(2^{-2/3}\tilde r,s) \sim a {\rm Ai}(- 2^{-1/3} s - \tilde r) + b {\rm Bi}(- 2^{-1/3} s - \tilde r) \;.
\end{eqnarray}
Under the (reasonable) assumption that $f(2^{-2/3}\tilde r,s)$ remains bounded, one obtains that $b = 0$. On the other hand, the amplitude $a$ can be fixed as we know the large $\tilde r$ behavior of $f(2^{-2/3}\tilde r,s)$ from Eq.~(\ref{fg_larger}). This yields $a = \sqrt{\pi} 2^{1/6}$ and, hence, from Eq. (\ref{schrod_alpha1/2}) together with (\ref{asympt_phi}) one obtains
\begin{eqnarray}\label{relation_Gf}
G(\tilde r,x) = 2^{-1/6} f(2^{-2/3} \tilde r,-2^{1/3} x) \;.
\end{eqnarray}

We now come back to the kernel $K_N(y-r,y-r')$ in the double scaling limit. From the above relation (\ref{relation_Gf}), together with the ${\bf B}$-equation of the Lax pair (\ref{lax_system}), one can express $\partial_x G(\tilde r,x)$ in terms of the functions $f$ and $g$ to obtain (\ref{kernel_dble_scaling}) in the double scaling limit [we recall that $y=\sqrt{2N}+\frac{x}{\sqrt{2}}N^{-\frac16}$ (\ref{double_scaling_NM})]
\begin{eqnarray}\label{kernel_double_1}
&&K_N\left(y - \frac{\tilde r}{\sqrt{2}}N^{-1/6}, y-\frac{\tilde r'}{\sqrt{2}}N^{-1/6}\right) \underset{N \to \infty}{\sim} \nonumber \\
&&N^{1/6} \sqrt{2}\frac{f(2^{-2/3} \tilde r,-2^{1/3}x) g(2^{-2/3}\tilde r',-2^{1/3}x)- g(2^{-2/3}\tilde r,-2^{1/3}x) f(2^{-2/3}\tilde r',-2^{1/3}x)}{\pi(\tilde r - \tilde r')} \;.
\end{eqnarray}
In view of this expression (\ref{kernel_double_1}), it is natural to introduce the following functions $\tilde f(\tilde r,x)$ and $\tilde g(\tilde r,x)$ such that
\begin{eqnarray}
f(2^{-2/3} \tilde r,-2^{1/3}x) = 2^{1/3} \tilde f(\tilde r,x) \;, \; g(2^{-2/3} \tilde r,-2^{1/3}x) = \tilde g(\tilde r,x) \;,
\end{eqnarray}
from which it follows, using Eq.~(\ref{relation_Gf}), that
\begin{eqnarray}\label{relation_Gftilde}
G(\tilde r,x) = 2^{1/6} \tilde f(\tilde r,x) \;.
\end{eqnarray}
Finally, one has
\begin{eqnarray}\label{KN_full}
K_N\left(y - \frac{\tilde r}{\sqrt{2}}N^{-1/6}, y-\frac{\tilde r'}{\sqrt{2}}N^{-1/6}\right) \underset{N \to \infty}{\sim} N^{1/6}2^{5/6} \frac{\tilde f(\tilde r,x) \tilde g(\tilde r',x) - \tilde f(\tilde r',x)\tilde g(\tilde r,x)}{\pi(\tilde r - \tilde r')} \;.
\end{eqnarray}
From the Lax pair (\ref{lax_system}), (\ref{def_AB}), together with some identities satisfied by $q_{1/2}(x)$, we show in Appendix~\ref{app_lax_pair} that $\tilde f(r,s)$ and $\tilde g(r,s)$ are solutions of a system of differential equations which can be expressed in terms of $q(x)$, the Hastings-McLeod solution of the Painlev\'e II equation with $\alpha =0$ (\ref{hasting}):
\begin{eqnarray}\label{lax_system_tilde}
\frac{\partial}{\partial \tilde r} 
\left(
\begin{array}{c}
\tilde f(\tilde r,x) \\
\tilde g(\tilde r,x) 
\end{array} \right) 
=  {\mathbf{\tilde A}}\left(
\begin{array}{c}
f(\tilde r,x) \\
g(\tilde r,x) 
\end{array} \right) \;, \;
 \frac{\partial}{\partial x} 
\left(
\begin{array}{c}
\tilde f(\tilde r,x) \\
\tilde g(\tilde r,x) 
\end{array} \right) 
=  {\mathbf{\tilde B}}\left(
\begin{array}{c}
\tilde f(\tilde r,x) \\
\tilde g(\tilde r,x) 
\end{array} \right) \;, \;
\end{eqnarray}
where ${\bf \tilde A}$ and ${\bf \tilde B}$ are $2 \times 2$ matrices given by
\begin{eqnarray}\label{def_AB_tilde}
{\bf \tilde A}
= 
\left( 
\begin{array}{cc}
-\frac{q'(x)}{q(x)} & 1+q^2(x)/\tilde r \\
-\tilde r-\frac{\int_x^\infty q^2(u)du}{q^2(x)} & \frac{q'(x)}{q(x)} 
\end{array}
\right) \;, \;
{\bf \tilde B}
= 
\left( 
\begin{array}{cc}
\frac{q'(x)}{q(x)} & -1 \\
\tilde r & - \frac{q'(x)}{q(x)} 
\end{array}
\right)\;, \;
\end{eqnarray}
where the solutions $\tilde f(\tilde r,x)$ and $\tilde g(\tilde r,x)$ are characterized by the asymptotic behavior inherited from~(\ref{fg_larger}) 
\begin{subequations}\label{fgtilde_larger}
\begin{eqnarray}
&&\tilde f(\tilde r,x) \underset{\tilde r \to \infty}{\sim} 2^{-1/6} \tilde r^{-1/4} \sin{\left(\frac{2}{3}\tilde r^{3/2} - x \sqrt{\tilde r} + \frac{\pi}{4} \right)} +Ê{\cal O}(\tilde r^{-3/4}) \;, \\
&&\tilde g(\tilde r,x) \underset{\tilde r \to \infty}{\sim} 2^{-1/6} \tilde r^{1/4} \cos{\left(\frac{2}{3}\tilde r^{3/2} - x \sqrt{\tilde r} + \frac{\pi}{4} \right)} +Ê{\cal O}(\tilde r^{-1/4}) \;.
\end{eqnarray}
\end{subequations}
From this expression (\ref{KN_full}) one can compute the DOS $\rho_{\rm DOS}(r,N)$ for $r \sim {\cal O}(N^{-1/6})$. The above asymptotics (\ref{fgtilde_larger}) will be useful to study the asymptotic behavior of $\rho_{\rm DOS}(r,N)$ for large $r$.

\subsubsection{Limiting expression of the DOS in the double scaling limit}

For the purpose of the computation of $\rho_{\DOS}(r,N)$ given in (\ref{rho_kernel}), we need to compute $K_N(y-{\tilde r}N^{-1/6}/{\sqrt{2}}, y)$ which is given, from Eq. (\ref{KN_full}), by
\begin{eqnarray}
&&K_N(y-r,y) \underset{N \to \infty}{\sim} N^{1/6} 2^{5/6}\frac{\tilde f(\tilde r,x) \tilde g(0,x)- \tilde g(\tilde r,x) \tilde f(0,x)}{\pi\tilde r} \;.
\end{eqnarray}
Using that $\tilde g(0,x) = 0$ (see  Appendix \ref{app_small_r}), together with $\tilde f(0,x) = 2^{-1/6} \sqrt{\pi} q(x)$ [see Eqs. (\ref{G_r0}), (\ref{relation_Gftilde})], one obtains:
\begin{eqnarray}
&&K_N\left(y-\frac{\tilde r}{\sqrt{2}}N^{-1/6},y\right) \underset{N \to \infty}{\sim} -N^{1/6} 2^{2/3} \frac{\tilde g(\tilde r,x) q(x)}{\sqrt{\pi} \tilde r} \;.
\end{eqnarray}

On the other hand, we also need the kernel at coinciding point (\ref{rho_kernel}). It can easily be obtained from Eq.~(\ref{KN_full}) using l'Hospital's rule
\begin{eqnarray}\label{limit_kernel_coinciding}
&&K_N(y - r, y-r) \underset{N \to \infty}{\sim} \frac{2^{5/6} N^{1/6}}{\pi}\left[ \partial_{\tilde r}\tilde f(\tilde r,x) \tilde g(\tilde r,x) - \tilde f(\tilde r,x) \partial_{\tilde r}\tilde g(\tilde r,x)\right] \\
&& = \frac{2^{5/6} N^{1/6}}{\pi} \left[\left(\tilde r + \frac{\int_x^\infty q^2(u)du}{q^2(x)}\right)\tilde f^2 - 2 \frac{q'(x)}{q(x)}\tilde f \tilde g + \left(1 + \frac{q^2(x)}{\tilde r}\right) \tilde g^2\right] \;.
\end{eqnarray}
In particular, using that $\tilde g(\tilde r,x) = {\cal O}(r)$, when $r \to 0$ (see Appendix \ref{app_small_r}) and $\tilde f(0,x) = 2^{-1/6} \sqrt{\pi} q(x)$ one finds from (\ref{limit_kernel_coinciding}) that
\begin{eqnarray}\label{check}
&&K_N(y, y) \underset{N \to \infty}{\sim} \sqrt{2} N^{1/6} \int_x^\infty q^2(u) du \;.
\end{eqnarray}
One can check that this relation (\ref{check}) is consistent with the identity shown above (\ref{identity_3}). Indeed, one has
\begin{eqnarray}
\log F_N(y) \underset{N \to \infty}{\to} \log {\cal F}_2(\sqrt{2}N^{1/6}(y-\sqrt{2N})) \;,
\end{eqnarray}
where ${\cal F}_2(x)$ is the Tracy-Widom distribution for $\beta = 2$, given by \cite{TW94a}
\begin{eqnarray}
\log {\cal F}_2(\sqrt{2}N^{1/6}(y-\sqrt{2N})) = - \int_x^\infty (u-x) q^2(u) du \;,
\end{eqnarray}
hence
\begin{eqnarray}\label{check2}
\frac{\partial}{\partial y} \log F_N(y) = \sqrt{2}N^{1/6}\int_x^\infty q^2(u) du \;.
\end{eqnarray}
One finds finally, that
\begin{subequations}\label{final_rho}
\begin{eqnarray}
&&\rho_{\DOS}(r,N) = N^{-5/6} \sqrt{2} \tilde \rho_{\rm edge}(\sqrt{2}N^{1/6} r) \;, \\
&& \tilde \rho_{\rm edge}(\tilde r) = \frac{2^{1/3}}{\pi} \int_{-\infty}^\infty \left[R \left(\left(\tilde r + \frac{R}{q^2}\right)\tilde f^2 - 2 \frac{q'}{q}\tilde f \tilde g + \left(1 + \frac{q^2}{\tilde r}\right) \tilde g^2\right)  - \frac{1}{\tilde r^2} q^2 \tilde g^2\right]  {\cal F}_2 \, dx \;,
\end{eqnarray}
\end{subequations}
where we have used the shorthand notations $q = q(x)$, $R=R(x)$, ${\cal F}_2 = {\cal F}_2(x)$, $\tilde f = \tilde f(\tilde r,x)$ and $\tilde g = \tilde g(\tilde r,x)$ as well as
\begin{eqnarray}\label{def_R}
R(x) = \int_x^\infty q^2(u) du \;.
\end{eqnarray}

This expression (\ref{final_rho}) is still a bit cumbersome but we can further simplify it, by using the remarkable identity
\begin{eqnarray}\label{amazing_id}
\partial_x \left(\left(\tilde r + \frac{R}{q^2}\right)\tilde f^2 - 2 \frac{q'}{q}\tilde f \tilde g + \left(1 + \frac{q^2}{\tilde r}\right) \tilde g^2\right) = -\tilde f^2 \;,
\end{eqnarray}
which can be checked using the fact that $\tilde f$ and $\tilde g$ are solutions of the Lax system (\ref{lax_system_tilde}). 
Hence using this identity in Eq. (\ref{amazing_id}) together with (\ref{def_R}), one can write $\tilde \rho_{\rm edge}(\tilde r)$ in Eq. (\ref{final_rho}) as
\begin{eqnarray}\label{rho_final_inter}
\tilde \rho_{\rm edge}(\tilde r) = \frac{2^{1/3}}{\pi} \int_{-\infty}^\infty \left[R \int_{x}^\infty \tilde f^2(\tilde r,u) \, du - q^2 \frac{\tilde g}{\tilde r^2}\right] {\cal F}_2(x) \, dx \;.
\end{eqnarray}
Besides, from the equation satisfied by $\partial_x \tilde g(\tilde r,x)$ (\ref{lax_system_tilde}, \ref{def_AB_tilde}), one has
\begin{eqnarray}\label{relation_fg}
\partial_x(q(x) \tilde g(\tilde r,x)) = \tilde r q(x)\tilde f(\tilde r,x) \Longrightarrow q(x) \tilde g(\tilde r,x) = -\tilde r \int_x^\infty q(u) \tilde f(\tilde r,u) \, du \;,
\end{eqnarray}
where we have used that $g(\tilde r,x) \to 0$ when $x \to \infty$. Performing an integration by part in (\ref{rho_final_inter}), and using that $R = {\cal F}_2'/{\cal F}_2$ together with (\ref{relation_fg}), we obtain
\begin{eqnarray}\label{rho_final_simplif}
\tilde \rho_{\rm edge}(\tilde r) = \frac{2^{1/3}}{\pi} \int_{-\infty}^\infty \left[\tilde f^2 - \frac{q^2}{\tilde r^2} \tilde g^2\right] {\cal F}_2 \, dx
= \frac{2^{1/3}}{\pi} \int_{-\infty}^\infty \left[\tilde f^2 - \left( \int_x^\infty q \tilde f du  \right)^2 \right] {\cal F}_2 \, dx \;,
\end{eqnarray}
where we have used the notation $q \tilde f = q(u) \tilde f(\tilde r,u)$. We recall that ${\cal F}_2(x)$ is the Tracy-Widom distribution associated to GUE, which is given by
\begin{eqnarray}
{\cal F}_2(x) = \exp{\left[-\int_x^\infty (u-x) q^2(u) du\right]} \;,
\end{eqnarray}
while $\tilde f(\tilde r,x)$ satisfies 
\begin{eqnarray}\label{schrod_text}
\partial_x^2 \tilde f(\tilde r,x) - [x + 2 q^2(x)] \tilde f(\tilde r,x) = - \tilde r \tilde f(\tilde r,x) \;, \; \tilde f(\tilde r,x) \underset{x \to \infty}{\sim} 2^{-1/6} \sqrt{\pi} {\rm Ai}(x-\tilde r) \;,
\end{eqnarray}
and $q(x)$ is the Hastings-Mc Leod solution of the Painlev\'e II equation (\ref{hasting}). The expression given in Eq. (\ref{rho_final_simplif}) is one of the main results of the present paper. We now analyze the asymptotic behaviors of $\tilde \rho_{\rm edge}(\tilde r)$ for small and large argument. 

\subsubsection{Asymptotic behavior for large $\tilde r$}

To analyze the large $\tilde r$ behavior of $\tilde \rho_{\rm edge}(\tilde r)$, it turns out that it is much easier to use the expression in Eq.~(\ref{final_rho}). To this purpose we use the large $\tilde r$ behavior of $\tilde f$ and $\tilde g$ in Eq.~(\ref{fgtilde_larger}). One finds that the leading terms in the integrand in (\ref{final_rho}) are
\begin{subequations}\label{expansion_larger}
\begin{eqnarray}
&&\left[R \left(\left(\tilde r + \frac{R}{q^2}\right)\tilde f^2 - 2 \frac{q'}{q}\tilde f \tilde g + \left(1 + \frac{q^2}{\tilde r}\right) \tilde g^2\right)  - \frac{1}{\tilde r^2} q^2 \tilde g^2\right] \sim R (\tilde r \tilde f^2 + \tilde g^2)  \\
&& \sim 2^{-1/3} \sqrt{\tilde r} R  \;.
\end{eqnarray}
\end{subequations}
Plugging this expansion (\ref{expansion_larger}) into the expression for $\tilde \rho_{\rm edge}(\tilde r)$ in Eq. (\ref{final_rho}) and using $R(x) {\cal F}_2(x)= d {\cal F}_2(x)/dx$ one obtains straightforwardly from (\ref{final_rho}) the leading asymptotic behavior
\begin{eqnarray}\label{asympt_rhotilde}
\tilde \rho_{\rm edge}(\tilde r) \sim \frac{\sqrt{\tilde r}}{\pi} \;,
\end{eqnarray}
which yields the second line of Eq. (\ref{asympt_tilderho}). As explained above, below Eq. (\ref{asympt_tilderho}), this asymptotic behavior (\ref{asympt_rhotilde}) ensures a perfect matching between the edge regime of $\rho_{\rm DOS}(r,N)$ when $r = {\cal O}(N^{-1/6})$, described by $\tilde \rho_{\rm edge}$ (\ref{final_rho}, \ref{rho_final_simplif}), and the bulk regime when $r = {\cal O}(\sqrt{N})$, described by a shifted Wigner semi-circle law in Eq. (\ref{eq:tilde_bulk}). Note that one can show, and this is confirmed by a numerical evaluation of our formula for $\tilde \rho_{\rm edge}(\tilde r)$ in Eq. (\ref{rho_final_simplif}) (see section \ref{section_numerical}), that the first corrections to the leading behavior in Eq. (\ref{asympt_rhotilde}) are of order ${\cal O}(\tilde r^{-1/2})$.

\subsubsection{Asymptotic behavior for small $\tilde r$}\label{section:large_r_DOS}

The study of the small $\tilde r$ behavior of $\tilde \rho_{\rm edge}(\tilde r)$, which is more conveniently performed on Eq. (\ref{rho_final_simplif}), requires the analysis of the Lax system (\ref{lax_system}) when $\tilde r \to 0$. Here we present the analysis of the small $\tilde r$ expansion at lowest order, which requires the expansion of $\tilde f(\tilde r,x)$ up to order ${\cal O}(\tilde r^3)$ (see Appendix~\ref{app_small_r}):
\begin{eqnarray}\label{smallr_exp_f}
\tilde f(\tilde r,x) = \tilde f(0,x) + \tilde r \tilde f_1(x) + \tilde r^2 \tilde f_2(x) + {\cal O}(r^3)\;,
\end{eqnarray} 
where the functions $\tilde f_k(x)$ can be computed explicitly along the lines explained in Appendix \ref{app_small_r} yielding:
\begin{subequations}\label{f0f1f2}
\begin{eqnarray}\
&&\tilde f(0,x) = 2^{-1/6} \sqrt{\pi} q \;, \label{f0}\\
&&\tilde f_1(x) = - 2^{-1/6} \sqrt{\pi}\left[q' + q R \right] \;, \label{f1} \\
&&\tilde f_2(x) = 2^{-{7}/{6}}\sqrt{\pi} \left[\frac{q'^2}{q} + q'R - \frac{R}{q} - \frac{1}{2}q^3 + \frac{q}{2} R^2 \right] \;.\label{f2}
\end{eqnarray}
\end{subequations}
One can then insert this expansion (\ref{smallr_exp_f}, \ref{f0f1f2}) into Eq. (\ref{rho_final_simplif}) to obtain
\begin{subequations}
\begin{eqnarray}\label{expansion_rho_small_r}
\tilde \rho_{\rm edge}(\tilde r) &=& \int_{-\infty}^\infty \left( q^2(x) - R^2(x) \right) {\cal F}_2(x) dx \label{term1} \\
&+& \tilde r \frac{2^{1/6}}{\sqrt{\pi}} \int_{-\infty}^\infty \left(q(x) \tilde f_1(x)  - R(x)\int_x^\infty q(u) \tilde f_1(u) du  \right)  {\cal F}_2(x) \, dx \; \label{term2} \\
&+& \tilde r^2 \Bigg[\frac{2^{1/6}}{\sqrt{\pi}}\int_{-\infty}^\infty \left(q(x) \tilde f_2(x)  - R(x)\int_x^\infty q(u) \tilde f_2(u) du  \right)  {\cal F}_2(x) \, dx \label{term3} \\
&+& \frac{2^{1/3}}{\pi}\int_{-\infty}^\infty \left(\tilde f_1^2(x) - \left(\int_x q(u) \tilde f_1(u)\right)^2\right) \Bigg] {\cal F}_2(x) \, dx + {\cal O}(r^3)\;. \label{term4}
\end{eqnarray}
\end{subequations}
By performing integration by parts, using $R(x) = {\cal F}_2'(x)/{\cal F}_2(x)$, one finds that the three first terms (\ref{term1})-(\ref{term3}) in this expansion 
all vanish. One finds finally
\begin{eqnarray}
\tilde \rho_{\rm edge}(\tilde r) &=& \tilde r^2  \int_{-\infty}^\infty \left[ (q' + q R)^2 - \frac{1}{4}(q^2- R^2)^2 \right] {\cal F}_2 \, dx + {\cal O}(\tilde r^3) \;.
\end{eqnarray}
After multiple integration by parts one can compute exactly this integral over $x$ (see Appendix \ref{app_small_r}). This yields the very simple result
\begin{eqnarray}\label{def_a2}
a_2 =  \int_{-\infty}^\infty \left[ (q' + q R)^2 - \frac{1}{4}(q^2- R^2)^2 \right] {\cal F}_2 \, dx = \frac{1}{2} \;,
\end{eqnarray}
so that, to leading order, $\tilde \rho_{\rm edge}(\tilde r) \sim \tilde r^2/2$. From the expression for $\tilde \rho_{\rm edge}(\tilde r)$ obtained in (\ref{rho_final_simplif}) it is possible to go beyond the leading order, as explained in Appendix \ref{app_small_r}. In fact, to compute $\tilde \rho_{\rm edge}(\tilde r)$ up to order ${\cal O}(\tilde r^4)$ one can show that this is sufficient to know $\tilde f(\tilde r,x)$ only up to order ${\cal O}(\tilde r^2)$, as given in Eq. (\ref{smallr_exp_f}). We then obtain explicitly that the term of order ${\cal O}(\tilde r^3)$ vanishes, yielding $\tilde \rho_{\rm edge}(\tilde r) \sim  \tilde r^2/2 + a_4 \tilde r^4$ where $a_4$ is given below (\ref{expr_a4}).

Finally, the asymptotic behaviors of $\tilde \rho_{\rm edge}(\tilde r)$ can be summarized, as announced in Eq. (\ref{asympt_tilderho}), as
\begin{eqnarray}
\tilde \rho_{\rm edge}(\tilde r) \sim
\begin{cases}\label{asympt_tilderho_text}
&\dfrac{1}{2} \, \tilde r^2  + a_4 \tilde r^4\;, \; \tilde r \to 0 \\
& \\
&\dfrac{\sqrt{\tilde r}}{\pi} \;, \; \tilde r \to \infty \\
\end{cases}\;,
\end{eqnarray}
where the amplitude $a_4$ is given by the integral
\begin{eqnarray}\label{expr_a4}
&&a_4 = \frac{1}{2} \int_{-\infty}^\infty \left[H(x) + \frac{1}{2} (T^2(x)- H^2(x)) \right] {\cal F}_2(x) \, dx \;, \\
&&H(x) = -\frac{1}{2} q^2 R + \frac{1}{6} R^3 + \int_x^\infty (q^4(u) + u q^2(u)) du \;, \; T(x) = \frac{H'(x)}{q(x)} \nonumber \;.
\end{eqnarray}
This integral (\ref{expr_a4}) can be evaluated numerically \cite{Praehofer_webpage} to yield $a_4 = -0.393575...$. The quadratic behavior of $\tilde \rho_{\rm edge}(\tilde r)$ for small $\tilde r$ can be qualitatively understood as it is related to the probability that two eigenvalues become extremely close to each other -- namely the first one and the second one [see Eq.~(\ref{id_gap_DOS})]. From the joint PDF of the eigenvalues (\ref{jPDF}), this probability vanishes quadratically for GUE. More generally, for the Gaussian $\beta$-ensemble, one thus expects that $\tilde \rho_{\rm edge}(\tilde r) \sim a_\beta \tilde r^\beta$, while there is no obvious simple argument to compute the constant $a_\beta$, even for $\beta = 2$ where $a_2 = 1/2$, which arises in our calculation in a non trivial way. On the other hand, if we assume a smooth matching between the edge region and the bulk region described by the shifted Wigner semi-circle law (\ref{shifted_wigner}), then the large $\tilde r$ behavior of $\tilde \rho_{\rm edge}(\tilde r)$, to leading order, immediately follows: $\tilde \rho_{\rm edge}(\tilde r) \sim {\sqrt{\tilde r}}/{\pi}$ to match with the small argument of Eq. (\ref{shifted_wigner}). It is natural to assume that the matching holds for all values of $\beta$. Hence one expects the following asymptotic behaviors 
\begin{eqnarray}
\tilde \rho_{\rm edge}(\tilde r) \sim
\begin{cases}\label{asympt_tilderho_beta}
&a_\beta \, \tilde r^\beta  + o(\tilde r^{\beta}) \;, \; \tilde r \to 0 \\
& \\
&\dfrac{\sqrt{\tilde r}}{\pi} + {o}(\tilde r^{1/2}) \;, \; \tilde r \to \infty \\
\end{cases}\;,
\end{eqnarray}
for the Gaussian $\beta$-ensemble.

We now show that the double scaling analysis carried out to compute $\tilde \rho_{\rm edge}(\tilde r)$ can be straightforwardly adapted to obtain a relatively simpler formula (compared to the formula obtained by WBF in Ref. \cite{WBF13} and given in Appendix \ref{formula_WBF}) for the distribution of the typical fluctuations of the first (scaled) gap in the large $N$ limit.

\section{Application to the distribution of the gap between the two largest eigenvalues}

We now focus on the first gap between the two largest eigenvalues. If one denotes by $\lmax = \Lambda_{1,N} \geq \Lambda_{2,N} \geq \cdots \geq \Lambda_{N,N}$ we compute the PDF $p_{\rm GAP}(r,N)$ of the random variable $d_{1,N} = \Lambda_{1,N} - \Lambda_{2,N}$ (see Fig. \ref{fig_illust}).

\subsection{Exact expression for any finite $N$ and large $N$ expansion}

An exact expression for $p_{\rm GAP}(r,N)$ can be easily written from the joint PDF of the $N$ eigenvalues in Eq.~(\ref{jPDF}). For this purpose, we notice that the probability $p_{\rm GAP}(r,N)dr = {\rm Pr.}(\Lambda_{1,N} - \Lambda_{2,N} \in [r,r+dr])$ is obtained by integrating over the value $y \in (-\infty, +\infty)$ of  the second eigenvalue $\Lambda_{2,N}$, while $\Lambda_{1,N} = \lmax$ is fixed to be $y+r$.  This means in particular that there are $(N-2)$ eigenvalues whose value is less than $y$. Hence:
\begin{equation}\label{rho_gap_first}
p_{\rm GAP}(r,N) = N (N-1) \int_{-\infty}^{+\infty} dy \int_{-\infty}^y d \lambda_1 \int_{-\infty}^y d \lambda_2 \cdots  \int_{-\infty}^y d \lambda_{N-2} P_{\rm joint}(\lambda_1, \cdots, \lambda_{N-2}, y, y +r) \;,
\end{equation}
 where the factor $N(N-1)$ in Eq. (\ref{rho_gap_first}) comes from the $N(N-1)$ ways of choosing the pair of the two largest eigenvalues $(\Lambda_{1,N}, \Lambda_{2,N})$ among $N$. By comparing with the exact expression in (\ref{rho_start}), using the invariance of the joint PDF (\ref{jPDF}) under permutation of two eigenvalues, one obtains that $p_{\rm GAP}(r,N) = (N-1) \rho_{\DOS}(-r,N)$, as announced in Eq. (\ref{id_gap_DOS}) \footnote{See the remark below Eq. (\ref{kernel_CD}) concerning the definition of $\rho_{\rm DOS}(r,N)$ for negative $r$.}. 

From this identity, we can now use the previous analysis to study the typical fluctuations of the first gap $d_{1,N}$, i.e. for $d_{1,N} = {\cal O}(N^{-1/6})$. This yields immediately the expression announced in Eq. (\ref{rho_intro_simplif}):
\begin{subequations}\label{expr_gap_text}
\begin{eqnarray}
&&p_{\rm GAP}(r,N) = \sqrt{2} N^{1/6} \tilde p_{\rm typ}\left( r\sqrt{2}N^{1/6} \right)  \;, \\
&&\tilde p_{\rm typ}(\tilde r) = \frac{2^{1/3}}{\pi} \int_{-\infty}^\infty \left[\tilde f^2(-\tilde r,x) - \left( \int_x^\infty q(u) \tilde f(-\tilde r,u) du  \right)^2 \right] {\cal F}_2(x) \, dx \;,
\end{eqnarray}
\end{subequations}
where $\tilde f(-\tilde r,x)$ is solution of the Schr\"odinger equation:
\begin{eqnarray}\label{schrod_intro}
\partial_x^2 \tilde f(-\tilde r,x) - [x + 2 q^2(x)] \tilde f(-\tilde r,x) =  \tilde r \tilde f(\tilde r,x) \;, \; \tilde f(-\tilde r,x) \underset{x \to \infty}{\sim} 2^{-1/6} \sqrt{\pi} {\rm Ai}(x+\tilde r) \;, \; \tilde r > 0 \;.
\end{eqnarray}
One can again show that $\tilde f(-\tilde r,x)$ is the first component of the $2d$-vector which is solution of the Lax system in Eq. (\ref{lax_system_tilde})
where the matrices $\tilde {\mathbf A}$ and  $\tilde {\mathbf B}$ are given in Eq. (\ref{def_AB_tilde}) with the substitution $\tilde r \to - \tilde r$. This change of sign has drastic consequences on the asymptotic behaviors of $\tilde f(-\tilde r,x)$ and $\tilde g(-\tilde r,x)$. Instead of an oscillating behavior as in Eq. (\ref{fgtilde_larger}) they are characterized in this case by an exponential decay. This is also clear on the above Schr\"odinger equation (\ref{schrod_intro}). One has indeed \cite{CK08}
\begin{subequations}\label{fgtilde_larger_gap}
\begin{eqnarray}
&&\tilde f(-\tilde r,s) \underset{\tilde r \to \infty}{\sim} 2^{-7/6} \tilde r^{-1/4} \exp{\left(-\frac{2}{3}\tilde r^{3/2} - s \sqrt{\tilde r}  \right)}\left( 1 +Ê{\cal O}(\tilde r^{-1/2}) \right) \;, \\
&&\tilde g(-\tilde r,s) \underset{\tilde r \to \infty}{\sim} 2^{-7/6} \tilde r^{1/4} \exp{\left(-\frac{2}{3}\tilde r^{3/2} - s \sqrt{\tilde r}  \right)}\left( 1 +Ê{\cal O}(\tilde r^{-1/2}) \right)  \;,
\end{eqnarray}
\end{subequations}
which are essential to study the large $\tilde r$ asymptotics of $\tilde p_{\rm typ}(\tilde r)$.  

\subsection{Asymptotic behavior of $\tilde p_{\rm typ}(\tilde r)$ for small argument}\label{section_smallarg}

The small $\tilde r$ asymptotic of $\tilde p_{\rm typ}(\tilde r)$ is exactly similar to the one performed before for $\tilde \rho_{\rm edge}(\tilde r)$ as one can argue that it involves only even powers of $\tilde r$. Indeed, for small $\tilde{r}$, the contribution to $\tilde{\rho}_{\rm edge}(\tilde{r})$ come only from the gap between the two largest eigenvalues because the others (the third, the fourth etc. eigenvalues) are too far. This implies that $\tilde \rho_{\rm edge}(\tilde r) \simeq \tilde p_{\rm typ}(\tilde r)$. But on the other hand the identity in (\ref{id_gap_DOS}) implies $\tilde p_{\rm typ}(\tilde r) \simeq \tilde \rho_{\rm edge}(-\tilde r)$. Hence, if one assumes that $\tilde \rho_{\rm edge}(\tilde r)$ is analytic in $\tilde r = 0$, one expects that the small $\tilde r$ expansion of $\tilde p_{\rm typ}(\tilde r)$ [or equivalently the one of $\tilde \rho_{\rm edge}(\tilde r)$] only involves even powers of $\tilde r$. In particular, the two first terms are [see Eq. (\ref{asympt_tilderho_text})]  
\begin{eqnarray}\label{gap_small_r}
\tilde p_{\rm typ}(\tilde r) = \frac{1}{2} \tilde r^2 + a_4 \tilde r^4 + {\cal O}(\tilde r^6) \;,
\end{eqnarray} 
as announced in the introduction in Eq. (\ref{asympt_tildegap}). The quadratic behavior in (\ref{gap_small_r}) can be understood physically as coming from the short distance repulsion between eigenvalues, which comes from the Vandermonde term $\prod_{i,j}(\lambda_i - \lambda_j)^2$ in the joint PDF  for GUE (\ref{jPDF}). Similarly to the DOS (\ref{asympt_tilderho_beta}), one thus expects that, for the Gaussian $\beta$-ensemble, one has $\tilde p_{\rm typ}(\tilde r) \sim a_\beta \tilde r^\beta$ (see also \cite{MG13} in the case of Gaussian Orthogonal Ensemble, corresponding to $\beta = 1$).

\subsection{Asymptotic behavior of $\tilde p_{\rm typ}(\tilde r)$ for large argument}

The asymptotic analysis of $\tilde p_{\rm typ}(\tilde r)$ for large $\tilde r$ is quite different from the one performed above for $\tilde \rho_{\rm edge}(\tilde r)$ in section \ref{section:large_r_DOS}. This difference, as mentioned above, is due to the qualitatively different behaviors of $\tilde f(\tilde r,x)$ and $\tilde g(\tilde r,x)$ in Eq.~(\ref{fgtilde_larger}) and $\tilde f(-\tilde r,x)$ and $\tilde g(-\tilde r,x)$ in Eq.~(\ref{fgtilde_larger_gap}) when $\tilde r$ is large. To perform this large $\tilde r$ analysis, it is more convenient to use the following expression in terms of $\tilde f(-\tilde r,x)$ and $\tilde g(-\tilde r,x)$ [as in the first expression given in Eq. (\ref{rho_final_simplif}) with the substitution $\tilde r \to - \tilde r$]:
\begin{eqnarray}\label{exp_ptyp}
\tilde p_{\rm typ}(\tilde r) = \frac{2^{1/3}}{\pi} \int_{-\infty}^\infty \left[\tilde f^2(-\tilde r,x) -\frac{1}{r^2} q^2(x) \tilde g^2(-\tilde r,x) \right] {\cal F}_2(x) \, dx \;.
\end{eqnarray}
By plugging the large $\tilde r$ behaviors of $\tilde f(-\tilde r,x)$ and $\tilde g(-\tilde r,x)$ given in Eq. (\ref{fgtilde_larger_gap}) in the above expression (\ref{exp_ptyp}), one obtains
\begin{eqnarray}\label{larger_step1}
\tilde p_{\rm typ}(\tilde r) = \frac{1}{4\pi} \frac{1}{\tilde r^{1/2}} e^{-\frac{4}{3} \tilde r^{3/2}} I(\tilde r) \left(1 + {\cal O}(\tilde r^{-1/2})\right) \;, \; I(\tilde r) = \int_{-\infty}^\infty e^{- 2 x \sqrt{\tilde r}}{\cal F}_2(x)  \, dx \;.
\end{eqnarray}
The leading corrections, of order ${\cal O}(\tilde r^{-1/2})$, in Eq. (\ref{larger_step1}) come from the leading corrections to the large $\tilde r$ behavior of $\tilde f(-\tilde r,x)$ in Eq. (\ref{fgtilde_larger_gap}) (these terms are analyzed in detail in Appendix \ref{app_large_r}). We thus focus now on the large $\tilde r$ expansion of $I(\tilde r)$ in (\ref{larger_step1}). To analyze it, we first split the integral over $x$ in (\ref{larger_step1}) into two parts: one on ${\mathbb R}^-$ and another one on ${\mathbb R}^+$:
\begin{subequations}\label{def_I+I_}
\begin{eqnarray}
&&I(\tilde r) = I_+(\tilde r) + I_-(\tilde r) \;, \\
&&I_+(\tilde r) = \int_{0}^\infty e^{- 2 x \sqrt{\tilde r}}{\cal F}_2(x)  \, dx \;, \; I_-(\tilde r) = \int_{-\infty}^0 e^{- 2 x \sqrt{\tilde r}}{\cal F}_2(x)  \, dx \;.
\end{eqnarray}
\end{subequations}
The analysis of $I_+(\tilde r)$ is easily done, using that ${\cal F}_2(x) \leq 1$, for all $x$, which implies
\begin{eqnarray}\label{bound_I+}
0 \leq I_+(\tilde r) \leq \int_0^\infty e^{-2x \sqrt{\tilde r}} \, dx = \frac{1}{2 \sqrt{\tilde r}} \;,
\end{eqnarray}  
such that $I_+(\tilde r) \to 0$ when $\tilde r \to \infty$. 

The analysis of $I_-(\tilde r)$ is quite different. Indeed, because of the exponential term $e^{-2 x \sqrt{\tilde r}}$, with $x < 0$, the integral over $x$ in $I_-(\tilde r)$ is dominated, when $\tilde r \gg 1$, by the region of large negative $x$. Hence for large $\tilde r$, one can replace, in $I_-(\tilde r)$, ${\cal F}_2(x)$ by its asymptotic behavior for large negative argument. It reads \cite{BBdF08}:
\begin{eqnarray}\label{asympt_TW}
{\cal F}_2(x) = \tau_2 |x|^{-1/8} e^{-\frac{1}{12} |x|^3} \left(1 + {\cal O}(|x|^{-3}) \right) \;, \; \tau_2 = 2^{1/24} e^{\zeta'(-1)} \;.
\end{eqnarray}
By plugging the asymptotic behavior of ${\cal F}_2(x)$ (\ref{asympt_TW}) into the integral defining $I_-(\tilde r)$ in Eq. (\ref{def_I+I_}) we see that the integrand is of the form $\exp{\left(2 |x| \sqrt{\tilde r} - \tfrac{1}{12}|x|^3\right)}$: hence for large $\tilde r$, this integral can be evaluated by the saddle point method. Besides, by balancing the two terms in the argument of the exponential $|x| \sqrt{\tilde r}$ and $|x|^3$, we obtain that the saddle point $x^*$ is of order ${\cal O}({\tilde r}^{1/4})$ -- and negative. This suggests to perform the change of variable $x = -u \tilde r^{1/4}$ in Eq.~(\ref{def_I+I_}):
\begin{subequations}\label{saddle}
\begin{eqnarray}
I_-(\tilde r) &=& \tilde r^{1/4} \int_0^\infty e^{2 u \tilde r^{3/4}} {\cal F}_2(- u \tilde r^{1/4}) \, du\\
&=& \tau_2 \tilde r^{7/32} \int_0^\infty \frac{du}{u^{1/8}} e^{\tilde r^{3/4} \phi(u)} \left(1 + {\cal O}(r^{-3/4}) \right) \;, \; \phi(u) = 2u - \frac{u^3}{12}\;, 
\end{eqnarray}
\end{subequations}
where the terms of order ${\cal O}(\tilde r^{-3/4})$ coming from the subleading corrections to ${\cal F}_2(x)$ in Eq. (\ref{asympt_TW}) are studied in detail in Appendix \ref{app_large_r}. It is easy to see that the function $\phi(u)$ in Eq. (\ref{saddle}) admits a maximum in $u^* = 2 \sqrt{2}$, where $\phi(u^*) = 8\sqrt{2}/3$ so that the integral over $u$ in (\ref{saddle}) can be evaluated using the saddle point method to yield
\begin{eqnarray}\label{gap_leading}
I_-(\tilde r) = {2^{5/48}}{\sqrt{\pi}} e^{\zeta'(-1)} \tilde r^{-\frac{5}{32}} \exp{\left(\frac{8}{3} \sqrt{2} \tilde r^{3/4} \right)}\left(1 +Ê{\cal O}(\tilde r^{-3/4}) \right) \;.
\end{eqnarray}
Using that $I(\tilde r) \sim I_-(\tilde r)$ [see Eqs. (\ref{def_I+I_}, \ref{bound_I+})] together with the analysis of the first corrections to this leading behavior (\ref{gap_leading}) which is performed in Appendix \ref{app_large_r}, we obtain finally the result :
\begin{eqnarray}\label{final_gap}
\tilde p_{\rm typ}(\tilde r) = A  \; \exp{\left(-\dfrac{4}{3}\tilde r^{3/2} + \dfrac{8}{3}\sqrt{2}\tilde r^{3/4}\right)}{\tilde r^{-{21}/{32}}} \left(1 - \dfrac{1405 \sqrt{2}}{1536} \tilde r^{-3/4} + {\cal O}(\tilde r^{-3/2})\right) \;,
\end{eqnarray} 
with $A = 2^{-91/48} e^{\zeta'(-1)}/\sqrt{\pi}$, as announced in the second line of Eq. (\ref{asympt_tildegap}). 

The asymptotic behavior of $\tilde p_{\rm typ}(\tilde r)$ to leading order, $\tilde p_{\rm typ}(\tilde r) \sim \exp{\left(-\frac{4}{3} \tilde r^{3/2}\right)}$ turns out to be the same as the right tail of the Tracy-Widom distribution, ${\cal F}_2$, in Eq. (\ref{def_TW2}). This can be understood via the following heuristic argument (see also \cite{WBF13}): for large separation $\tilde r$ the two first eigenvalues $\Lambda_{2,N}$ and $\Lambda_{1,N} = \lmax$ become statistically independent:
\begin{eqnarray}\label{heuristic}
\tilde p_{\rm gap}(\tilde r) &=& \int_{-\infty}^\infty {\rm Pr.} [\Lambda_{2,N} = \lambda, \lmax = \lambda+\tilde r] d\lambda \underset{r \to \infty}{\sim} \int_{-\infty}^\infty {\rm Pr.} [\Lambda_{2,N} = \lambda] {\rm Pr.} [\lmax = \lambda+\tilde r] d\lambda \nonumber \\
&\underset{r \to \infty}{\sim}& {\rm Pr.} [\lmax = \tilde r] \;,
\end{eqnarray}
which thus naturally yields the right tail of the TW distribution. Note this heuristic argument (\ref{heuristic}) can be adapted to the Gaussian $\beta$-ensemble, for which the TW-$\beta$ distribution behaves for large argument like ${\cal F}'_\beta(x) \sim \exp{\left(-\frac{2\beta}{3}x^{3/2}\right)}$. Hence, reminding the small $\tilde r$ behavior of $\tilde \rho_{\rm edge}(\tilde r)$ studied in section \ref{section_smallarg}, one expects, for the 
Gaussian $\beta$-ensemble:
\begin{eqnarray}\label{asympt_tildegap_beta}
\tilde p_{\rm typ}(\tilde r) \sim
\begin{cases}
a_{\beta} \tilde r^\beta  + o(\tilde r^\beta) & \tilde r \to 0 \\
\\
\exp{\left(-\dfrac{2\beta}{3}\tilde r^{3/2} + o(\tilde r^{3/2}) \right)} \;, \; & \tilde r \to +\infty \;,
\end{cases}
\end{eqnarray} 
while the computation of $a_{\beta}$ as well as the subleading corrections, both for small and large arguments, for any $\beta$ remains challenging.

\section{Numerical evaluation of the limiting scaling functions $\tilde \rho_{\rm edge}(\tilde r)$ and $\tilde p_{\rm typ}(\tilde r)$}\label{section_numerical}

In this section, we provide a numerical evaluation of the limiting functions $\tilde \rho_{\rm edge}(\tilde r)$ and $\tilde p_{\rm typ}(\tilde r)$ given respectively in Eq. (\ref{rho_final_simplif}) and Eq. (\ref{expr_gap_text}). To evaluate numerically these formula, we need to compute numerically the solution of the Lax pair, $(\tilde{f},\tilde{g})$. We found that the easiest way to evaluate them is to solve numerically (with Mathematica) the following system of coupled differential equations satisfied by $\tilde f(\tilde r,x)$ (\ref{schrod_text}) and $\tilde g(\tilde r,x)$ (\ref{relation_fg}), $\tilde{r}$ being a parameter: 
\begin{eqnarray}
\left\{
     \begin{array}{lll}
			{\partial_x^2}{} \tilde{f}(\tilde{r},x) = (x+2q^2-\tilde{r})\tilde{f}(\tilde{r},x) &\,,\,& \tilde{f}(\tilde{r},x) \underset{x\gg 1}{\approx} 2^{-1/6} \sqrt{\pi} {\rm Ai}(x-\tilde{r}) \;, \\
			 \partial_x \left[ q(x)\tilde{g}(\tilde{r},x)\right] = \tilde{r} q(x)\tilde{f}(\tilde{r},x) &\,,\,& \tilde{g}(\tilde{r},x) \underset{x\gg 1}{\to} 0 \;,\\
     \end{array}
     \right.
     \label{system_eq_numerique}
\end{eqnarray}
where $q$ is the Hasting McLeod solution of Painlev\'e II (\ref{hasting}) whose numerical estimation is based on M.~Pr\"ahofer's tables \cite{Praehofer_webpage}. Once $\tilde f(\tilde r,x)$ and $\tilde g(\tilde r,x)$ are evaluated numerically, we can then compute the distributions $\tilde \rho_{\rm edge}(\tilde r)$ and $\tilde p_{\rm typ}(\tilde r)$ by evaluating numerically the integrals in Eq. (\ref{rho_final_simplif}) and Eq. (\ref{expr_gap_text}). 

\begin{figure}
\begin{center}
\psfrag{rhoedge}[c][][3]{$\tilde{\rho}_{\rm edge}(\tilde{r})$}
\psfrag{r}[c][][3]{$\tilde{r}$}
\psfrag{0}[c][][2.5]{$0$}
\psfrag{10}[c][][2.5]{$10$}
\psfrag{1}[c][][2.5]{$1$}
\rotatebox{-90}{\resizebox{70mm}{!}{\includegraphics{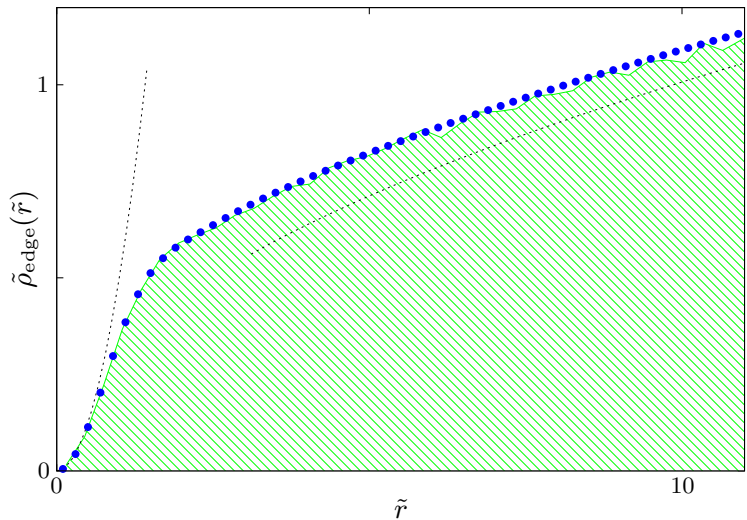}}}
\caption{Blue dots: numerical evaluation of our formula~(\ref{rho_final_simplif}) for $\tilde{\rho}_{\rm edge}(\tilde{r})$, green filling: numerical result with $2 \times 10^5$ realizations for a $1000\times1000$ GUE matrix,
black dashed curve: asymptotic behaviors given in Eq. (\ref{asympt_tilderho}).}
\label{Evaluation_numerique_rho}
\end{center}
\end{figure}
In Fig. \ref{Evaluation_numerique_rho} we show a plot of the scaling function $\tilde \rho_{\rm edge}(\tilde r)$, evaluated numerically (the blue dots), as a function of $\tilde r$. The black dashed curves indicate the leading behavior for small $\tilde r$, $\tilde \rho_{\rm edge}(\tilde r) \sim \tilde r^2/2$ [see Eq. (\ref{asympt_tilderho_text})] and the leading large $\tilde r$ behavior, $\tilde \rho_{\rm edge}(\tilde r) \sim \sqrt{\tilde{r}}/\pi$. A numerical estimation of $\tilde \rho_{\rm edge}(\tilde r) - \sqrt{\tilde{r}}/\pi$ indicates that this quantity vanishes like $\tilde r^{-1/2}$ for large $\tilde r$. Finally, the green filling in Fig. \ref{Evaluation_numerique_rho} shows a numerical estimation of $\tilde \rho_{\rm edge}(\tilde r)$ obtained by sampling $2 \times 10^5$ independent random Hermitian matrices of size $1000\times1000$. The agreement with the numerical evaluation of our formula (\ref{rho_final_simplif}) is quite good, which provides a good check of this formula (\ref{rho_final_simplif}). One should however notice that we have not attempted to evaluate the precision of our numerical procedure to evaluate numerically this formula (\ref{rho_final_simplif}).

\begin{figure}
\begin{center}
\psfrag{pgap}[c][][3]{$\tilde{p}_{\rm typ}(\tilde{r})$}
\psfrag{r}[c][][3]{$\tilde{r}$}
\psfrag{0}[c][][2.5]{$0$}
\psfrag{1}[c][][2.5]{$1$}
\psfrag{5}[c][][2.5]{$5$}
\psfrag{0.5}[c][][2.5]{$0.5$}
\rotatebox{-90}{\resizebox{70mm}{!}{\includegraphics{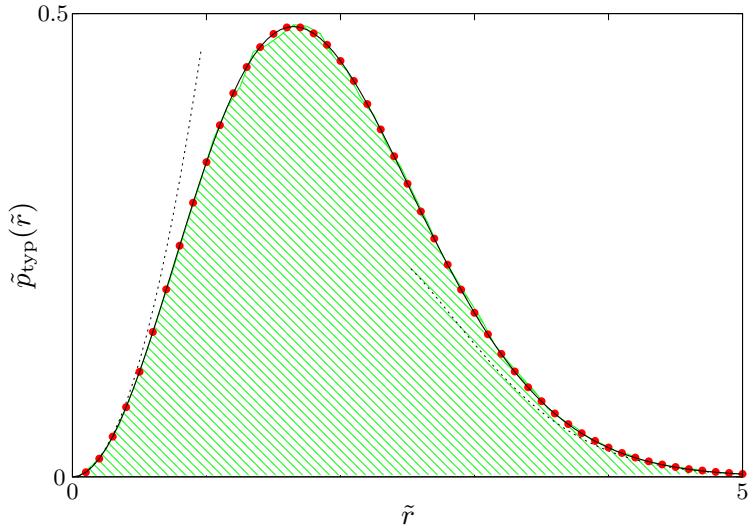}}}
\caption{Red dots: evaluation of our formula (\ref{expr_gap_text}) for $\tilde{p}_{\rm typ}(\tilde{r})$, 
black solid curve: WBF numerical result,
green filling: numerical result with $2 \times 10^5$ realizations for a $1000\times1000$ GUE matrix,
black dashed curve: asymptotic behaviors given in Eq. (\ref{asympt_tildegap}).}
\label{Evaluation_numerique_rho12}
\end{center}
\end{figure}

We have also evaluated numerically our formula for $\tilde p_{\rm typ}(\tilde r)$, given in Eq. (\ref{expr_gap_text}). These numerical values correspond to the red dots shown in Fig. \ref{Evaluation_numerique_rho12}. We have compared these values with the numerical evaluation of $\tilde p_{\rm typ}(\tilde r)$ provided by WBF \cite{WBF13} (the black curve in Fig. \ref{Evaluation_numerique_rho12}) -- this PDF is actually tabulated in Table 2 of Ref. \cite{WBF13}. Note that these numerical estimates have been obtained by evaluating numerically the expression of $\tilde p_{\rm typ}(\tilde r)$ in terms of Fredholm determinant \cite{TW94a} and not from a direct evaluation of their formula in Eqs. (\ref{WBF_1}), (\ref{WBF_2}), which we recall in Appendix \ref{formula_WBF}, in terms of an isomonodromic system associated to a Painlev\'e II transcendent. As explained in Ref. \cite{WBF13}, this formula (\ref{WBF_1}), (\ref{WBF_2}), at variance with the one in terms of Fredholm determinant \cite{WBF13}, can not be determined to arbitrary accuracy. This comparison shows a good agreement between the numerical evaluation of our formula in Eq. (\ref{expr_gap_text}) and the values tabulated by WBF in \cite{WBF13}. In Fig. \ref{Evaluation_numerique_rho12} we have also shown the data for $\tilde p_{\rm typ}(\tilde r)$ which we have 
computed numerically by sampling  $2 \times 10^5$ independent Hermitian matrices of size $1000\times1000$. The results of this numerical computation is shown in green filling in Fig.~\ref{Evaluation_numerique_rho12} and shows a nice agreement with both formula. 
Finally, in Fig. \ref{Evaluation_numerique_rho12}, we have also plotted the asymptotic behaviors (\ref{asympt_tildegap}) which we have obtained for
small $\tilde r$ for large $\tilde r$ (black dashed curve). A more precise analysis of the small and large $\tilde r$ of the data of WBF \cite{WBF13} shows actually a very good agreement with our precise asymptotics (\ref{asympt_tildegap}).

\section{Conclusion and perspectives}

To conclude, we have studied two different statistical characterizations of near-extreme eigenvalues of random matrices 
belonging to GUE: (i) the mean density of states $\rho_{\rm DOS}(r,N)$ which is the mean density of eigenvalues located
at a distance $r$ from the largest one, $\lmax$ and (ii) the PDF $p_{\rm GAP}(r,N)$ of the first gap, i.e. the spacing between the two largest
eigenvalues. The DOS has recently been studied in the context of i.i.d. random variables \cite{SM2007} and more recently for Brownian motion \cite{PCMS13}. It is a natural object to characterize the important phenomenon of ``crowding'' near extremes. The study presented in this paper is a first
attempt to characterize the ``crowding'' near the largest eigenvalue, $\lmax$, of random matrices. Here, we have performed a detailed
study of the DOS for GUE and it would be interesting to address the same questions for Gaussian $\beta$-ensembles, for which we could
only predict heuristically some asymptotic behaviors (\ref{asympt_tildegap_beta}). It would also be interesting to study the density of states close to the smallest eigenvalue $\lambda_{\rm min}$ at the hard edge of Laguerre-Wishart matrices, where we would expect quite different behaviors. 

We showed that these two quantities are actually related via the relation $p_{\rm GAP}(r,N) = (N-1) \rho_{\rm DOS}(-r,N)$ and
showed that both quantities can be conveniently expressed in terms of orthogonal polynomials which can be viewed as a generalization
of Hermite polynomials defined on a semi-infinite interval (\ref{def_op_intro}). We showed, using a double scaling analysis, that these
OP are related to the Lax pair of a Painlev\'e XXXIV equation, a result which can also be obtained using rather sophisticated method relying on Riemann-Hilbert techniques \cite{CK08}. 

We could then characterize the PDF characterizing the typical fluctuations $\tilde p_{\rm typ}(\tilde r)$ of the first gap, which are of order ${\cal O}(N^{-1/6})$, in terms of the solution of this Lax pair. Using this expression, we could also derive precise asymptotics of this PDF (\ref{asympt_tildegap}). Our expression is different, and also simpler, than the one found previously by Witte, Bornemann and Forrester \cite{WBF13}, and it would be very interesting to show explicitly that these two formulas do coincide. Note that similar relations arising in related contexts have recently emerged in \cite{Sch12,BLS12,Del12}. It would also be natural to extend the present study of the first gap to other ensembles of RMT, and in particular to the Gaussian $\beta$-ensembles.  

Finally, we have studied here the {\it typical} fluctuations of the first spacing at the edge, which are of order ${\cal O}(N^{-1/6})$. An interesting continuation of the present work would be, as it has been done in detail for the largest eigenvalue $\lmax$ itself (for a recent review see \cite{MS13}), to study the {\it atypically} large fluctuations of this first spacing, when they are of order ${\cal O}(N^{1/2})$. We leave these interesting open questions for future investigations.

\acknowledgement

We would like to thank S. N. Majumdar for very stimulating discussions and a careful reading of our manuscript. We acknowledge support by ANR grant 2011-BS04-013-01 WALKMAT. G. S. also acknowledges support from Labex PALM (Project RANDMAT).

%\newpage 

\appendix

\section{Rewriting the Lax system in terms of rescaled variables}\label{app_lax_pair}

In this appendix, we derive two identities relating the Hastings-McLeod solution of Painlev\'e II with $\alpha = 0$, denoted by $q(x)$ (\ref{hasting}), and the one with $\alpha =1/2$, denoted by $q_{1/2}$ (\ref{P2_alpha1/2}). We start with the relation between $q(x)$ and $q_{1/2}(x)$ \cite{WBF13,CJP99}
\begin{eqnarray}\label{Eq1_app1}
q_{1/2}(x) = -2^{-1/3} \frac{q'(-2^{-1/3}x)}{q(-2^{-1/3}x)} \;.
\end{eqnarray}
By taking the derivative with respect to (w.r.t.) $x$ on both sides of Eq. (\ref{Eq1_app1}), one obtains
\begin{eqnarray}\label{Eq2_app1}
q_{1/2}'(x) = 2^{-2/3} \frac{q''(-2^{-1/3}x)}{q(-2^{-1/3}x)} - 2^{-2/3} \left( \frac{q'(-2^{-1/3}x)}{q(-2^{-1/3}x)}\right)^2 \;.
\end{eqnarray}
By combining Eq. (\ref{Eq1_app1}) and Eq. (\ref{Eq2_app1}), one obtains
\begin{eqnarray}
x + 2 q_{1/2}^2(x) + 2 q_{1/2}'(x) = 2^{1/3} \left(2^{-1/3}x + \frac{q''(-2^{-1/3}x)}{q(-2^{-1/3}x)} \right) \;.
\end{eqnarray}
Finally, using that $q(x)$ is solution of the Painlev\'e II equation (\ref{hasting}) one obtains 
\begin{eqnarray}
q_{1/2}^2(x) + q_{1/2}'(x) + \frac{x}{2} = 2^{1/3} q^2(-2^{-1/3}x) \;. \label{identite_1}
\end{eqnarray}

We now derive a second identity by considering the following function
\begin{eqnarray}\label{def_J}
J(x) = - x - 2 q_{1/2}^2(x) + 2 q_{1/2}'(x)  \;.
\end{eqnarray}
It is straightforward to check that $J(x)$ satisfies
\begin{eqnarray}\label{ed_J}
J'(x) = - 2 q_{1/2}(x) J(x) - 2 \;.
\end{eqnarray}
The solution of the homogenous equation, $J'(x) = - 2 q_{1/2}(x) J(x)$ is, using (\ref{Eq1_app1}), $J(x) = A/q^2(-2^{-1/3}x)$. By varying the constant, one finds the solution of (\ref{ed_J}) under the form:
\begin{eqnarray}\label{constant_a}
J(x) = \frac{A(x)}{q^{2}(-2^{-1/3})x} \;, \; A(x) = 2^{4/3} \int_{-2^{-1/3}x}^\infty q^2(u) du \; +  \; a \;, 
\end{eqnarray}
where $a$ is a constant, independent of $x$. On the other hand, from the large $x$ behavior $q_{1/2}(x) \sim 1/(2x)$, one sees that $J(x)$ in (\ref{def_J}) behaves like $J(x) \sim -x$, when $x \to \infty$. This implies that the constant $a$ in Eq. (\ref{constant_a}) is $a=0$. Hence, one obtains a second identity:
\begin{eqnarray}\label{identite_2}
&&- q_{1/2}^2(x) + q_{1/2}'(x) -\frac{x}{2} = -2^{1/3} \frac{\int_{-2^{-1/3}x}^\infty q^2(u) du}{q^2(-2^{-1/3}x)} \;.
\end{eqnarray}
One can then use these identities (\ref{identite_1}) and (\ref{identite_2}) to write the matrix elements of the matrices ${\bf A}$ and ${\bf B}$ in Eq.~(\ref{def_AB}) in terms of $q(x)$ only -- and not $q_{1/2}(x)$. This yields ultimately, with an appropriate change of variables, the expression of the matrices ${\bf \tilde A}$ and ${\bf \tilde B}$ in Eq. (\ref{def_AB_tilde}).

\section{Expansion of the solution of the Lax system for small $\tilde r$}\label{app_small_r}

In this appendix, we give some details concerning the expansion of $\tilde \rho_{\rm edge}(\tilde r)$ [which is actually similar to the one of $\tilde p_{\rm typ}(\tilde r)$, see Eq. (\ref{id_gap_DOS})] for small argument $\tilde r$.

\subsection{General structure of the psi-functions $\tilde f(\tilde r,x)$ and $\tilde g(\tilde r,x)$ at small $\tilde r$}

The small $\tilde r$ expansion of $\tilde \rho_{\rm edge}(\tilde r)$ necessitates the expansion of the solution of the Lax pair $\tilde f(\tilde r,x)$ and $\tilde g(\tilde r,x)$, which we recall are solutions of the system of differential equations
\begin{eqnarray}\label{lax_system_tilde_app}
\frac{\partial}{\partial \tilde r} 
\left(
\begin{array}{c}
\tilde f(\tilde r,s) \\
\tilde g(\tilde r,s) 
\end{array} \right) 
=  {\mathbf{\tilde A}}\left(
\begin{array}{c}
f(\tilde r,s) \\
g(\tilde r,s) 
\end{array} \right) \;, \;
 \frac{\partial}{\partial s} 
\left(
\begin{array}{c}
\tilde f(\tilde r,s) \\
\tilde g(\tilde r,s) 
\end{array} \right) 
=  {\mathbf{\tilde B}}\left(
\begin{array}{c}
\tilde f(\tilde r,s) \\
\tilde g(\tilde r,s) 
\end{array} \right) \;, \;
\end{eqnarray}
where ${\bf \tilde A}$ and ${\bf \tilde B}$ are $2 \times 2$ matrices given by
\begin{eqnarray}\label{def_AB_tilde_app}
{\bf \tilde A}
= 
\left( 
\begin{array}{cc}
-\frac{q'(s)}{q(s)} & 1+q^2(s)/\tilde r \\
-\tilde r-\frac{\int_s^\infty q^2(u)du}{q^2(s)} & \frac{q'(s)}{q(s)} 
\end{array}
\right) \;, \;
{\bf \tilde B}
= 
\left( 
\begin{array}{cc}
\frac{q'(s)}{q(s)} & -1 \\
\tilde r & - \frac{q'(s)}{q(s)} 
\end{array}
\right)\;, \;
\end{eqnarray}
where the solutions $\tilde f(r,s)$ and $\tilde g(r,s)$ are characterized by the asymptotic behaviors given in Eq. (\ref{fgtilde_larger}) in the text. We have already shown that $\tilde f(\tilde r =0,x)$ exists, and in fact $\tilde f(\tilde r =0,x) = 2^{-1/6} \sqrt{\pi} q(x)$ and given the $\tilde r$-dependence of the matrices ${\bf \tilde A}$ and ${\bf \tilde B}$ one expects that $\tilde f(\tilde r,x)$ admits the following expansion
\begin{eqnarray}\label{f_small_r_gen}
\tilde f(\tilde r,x) = 2^{-1/6} \sqrt{\pi} q(x) + \sum_{n = 1}^\infty \tilde r^n \tilde f_n(x) \;.
\end{eqnarray}
To obtain the small $\tilde r$ expansion of $\tilde g(\tilde r,x)$, we show that it can be actually expressed in terms of $\tilde f(\tilde r,x)$. Using the relation (\ref{relation_fg}) shown in the text, one obtains from (\ref{f_small_r_gen}) that $\tilde g(\tilde r,x)$ admits the following expansion
\begin{eqnarray}\label{g_small_r_gen}
\tilde g(\tilde r,x) = \sum_{n=1}^\infty \tilde r^n \tilde g_n(x) \;, \; \tilde g_n(x) = -\frac{1}{q(x)} \int_x^\infty q(u) \tilde f_{n-1}(u) \, du \;, \; n \geq 1 \;.
\end{eqnarray}
By injecting the expansion of $\tilde f(\tilde r,x)$ (\ref{f_small_r_gen}) into the '$\tilde {\mathbf B}$-equation' satisfied by $\tilde f(\tilde r,x)$ 
[see Eqs. (\ref{lax_system_tilde_app}), (\ref{def_AB_tilde_app})] one finds the following equations
\begin{eqnarray}\label{recurrence}
&&\partial_x \tilde f_k(x) = \frac{q'(x)}{q(x)} \tilde f_k(x) - \tilde g_k(x) \;,
\end{eqnarray}
where $\tilde g_k(x)$ defined in Eq. (\ref{g_small_r_gen}) can be expressed in terms of $\tilde f_{k-1}(x)$. Hence this set of equations (\ref{recurrence}) can be solved iteratively for successive values of $k$, starting from $k=1$, to yield the first functions given in Eq.~(\ref{f0f1f2}) in the text.

\subsection{Lowest order expansion: calculation of the integral $a_2$}

In this section of the appendix, we show that the amplitude $a_2$ defined through the rather complicated integral [see Eq. (\ref{def_a2}) in the text] 
\begin{eqnarray}\label{def_a2_app}
a_2 =  \int_{-\infty}^\infty \left[ (q' + q R)^2 - \frac{1}{4}(q^2- R^2)^2 \right] {\cal F}_2 dx  \;,
\end{eqnarray}
has actually a very simple expression, namely $a_2 = 1/2$. First we recall that
\begin{eqnarray}\label{RF2}
R(x) = \int_{x}^\infty q^2(u) \, du = \frac{{\cal F}'_2(x)}{{\cal F}_2(x)} \;.
\end{eqnarray}
This identity (\ref{RF2}) is the crucial one as it allows us to compute this integral in (\ref{def_a2_app}), by using successive integration by parts. To this purpose, we first expand the squares in the integrand in (\ref{def_a2_app}) and decompose it~as
\begin{subequations}
\begin{eqnarray}
&&a_2 = \int_{-\infty}^\infty \left(q'^2 - \frac{1}{4}q^4 \right) {\cal F}_2 \, dx + J_1 + J_2 + J_3 \;, \label{decomp_J123} \\
&&J_1 = 2 \int_{-\infty}^\infty q q' R  {\cal F}_2 \, dx = 2 \int_{-\infty}^\infty q q' {\cal F}'_2 \, dx \;, \label{J1} \\
&&J_2 = \frac{3}{2} \int_{-\infty}^\infty  q^2 R^2 {\cal F}_2 \, dx = \frac{3}{2} \int_{-\infty}^\infty q^2 R {\cal F}'_2 \, dx \;, \label{J2} \\
&&J_3 = - \frac{1}{4} \int_{-\infty}^\infty R^4 {\cal F}_2 \, dx = - \frac{1}{4} \int_{-\infty}^\infty R^3 {\cal F}'_2 \, dx \;, \label{J3}
\end{eqnarray}
\end{subequations}
where we used the shorthand notations $q\equiv q(x), R\equiv R(x)$ and ${\cal F}_2 \equiv {\cal F}_2(x)$. We compute the integral $J_1$ in (\ref{J1}) by using an integration by part [integrating ${\cal F}_2'(x)$], which yields 
\begin{eqnarray}\label{final_J1}
J_1 = - 2 \int_{-\infty}^\infty \left(q'^2 + 2 q^4 + x q^2\right) {\cal F}_2 \, dx \;,
\end{eqnarray}
where we have used that $q(x)$ is solution of the Painlev\'e II equation (\ref{hasting}) as well as the asymptotic behavior of ${\cal F}_2(x)$ for large negative argument (\ref{asympt_TW}). Similarly, the integral $J_2$ in (\ref{J2}) can be transformed by using a similar integration by part [again integrating ${\cal F}_2'(x)$]. This yields
\begin{eqnarray}\label{final_J2}
J_2 = \frac{3}{2} \int_{-\infty}^\infty q^4 {\cal F}_2 \, dx - \frac{3}{2} J_1 = \int_{-\infty}^\infty \left(3q'^2 + \frac{15}{2} q^4 + 3 x q^2 \right) {\cal F}_2 \, dx \;.
\end{eqnarray}
The integral $J_3$ in (\ref{J3}) can be transformed using the same procedure as
\begin{eqnarray}\label{final_J3}
J_3 = - \frac{J_2}{2} = -\frac{1}{2} \int_{-\infty}^\infty \left(3q'^2 + \frac{15}{2} q^4 + 3 x q^2 \right) {\cal F}_2 \, dx \;.
\end{eqnarray}
Combining Eqs. (\ref{decomp_J123})- (\ref{final_J3}) one obtains
\begin{eqnarray}\label{a2_last_step}
a_2 = \frac{1}{2} \int_{-\infty}^\infty \left(q'^2 -q^4 - xq^2 \right) {\cal F}_2 \, dx \;.
\end{eqnarray}
Finally, this last integral can be computed exactly by using the identity:
\begin{eqnarray}\label{identite_R}
R(x) = \int_x^\infty q^2(u) \, du = q'^2 - q^4 - xq^2 \;,
\end{eqnarray}
which can be checked easily by taking derivative with respect to $x$ on both sides, and using that $q(x)$ is solution of the Painlev\'e II equation (\ref{hasting}). Hence using this identity (\ref{identite_R}), $a_2$ in (\ref{a2_last_step}) can be computed as
\begin{eqnarray}
a_2 = \frac{1}{2} \int_{-\infty}^\infty R(x) {\cal F}_2(x) \, dx = \frac{1}{2} \int_{-\infty}^\infty {\cal F}'_2(x) \, dx = \frac{1}{2} \;,
\end{eqnarray}
as given in Eq. (\ref{def_a2}) in the text.

\section{Large $\tilde r$ expansion of $\tilde p_{\rm typ}(\tilde r)$: beyond the leading order}\label{app_large_r}

In this appendix, we analyze in detail the large $\tilde r$ asymptotic of $\tilde p_{\rm typ}(\tilde r)$. We obtain in particular the first sub-leading corrections to the leading term obtained in the text, in Eq. (\ref{gap_leading}), yielding the rather precise asymptotics for  $\tilde p_{\rm typ}(\tilde r)$ 
given in Eq. (\ref{final_gap}). 

This expansion, beyond leading order, requires the determination of the first correction, of order ${\cal O}(\tilde r^{-1/2})$ to the asymptotic behavior of $\tilde f(-\tilde r,x)$, for large $\tilde r$, given in Eq. (\ref{fgtilde_larger_gap}), which we first focus on. We expect, from (\ref{fgtilde_larger_gap}), the following asymptotic behavior valid for large $\tilde r$:
\begin{eqnarray}\label{exp_ftilde_app}
\tilde f(-\tilde r, x) = \frac{1}{2^{7/6}} \tilde r^{-1/4} e^{-\frac{2}{3}\tilde r^{2/3} - x \sqrt{\tilde r}} \left(1 + \frac{1}{\sqrt{\tilde r}} F_1(x) + o(\tilde r^{-1/2})  \right) \;.
\end{eqnarray}
To compute the function $F_1(x)$, we use that $\tilde f(-\tilde r,x)$ satisfies the following Schr\"odinger equation
\begin{eqnarray}\label{schrod_app}
\partial_x^2 \tilde f(-\tilde r,x) - (x+2q^2(x))\tilde f(-\tilde r,x) = \tilde r \tilde f(-\tilde r,x) \;.
\end{eqnarray}
By injecting the asymptotic expansion (\ref{exp_ftilde_app}) into Eq. (\ref{schrod_app}), one obtains $F_1(x)$ as
\begin{eqnarray}\label{def_F}
F_1(x) = -\frac{1}{2} \int_{-\infty}^x \left[u + 2q^2(u)\right] \, du \;,
\end{eqnarray}
where we have used that $\lim_{x \to -\infty}F_1(x) = 0$. Although this is a reasonable assumption we have not been able to establish it rigorously. In the following, we will need the behavior of $F_1(x)$ for large negative argument. It can be obtained from the large negative argument of $q(s)$:
\begin{eqnarray}
q(s) = \sqrt{-\frac{s}{2}} \left( 1 + \frac{1}{8s^3} + {\cal O}(s^{-6})\right) \;, \; {\rm when \;} s \to -\infty \;,
\end{eqnarray}
which yields, for $F_1(x)$ in (\ref{def_F})
\begin{eqnarray}\label{asympt_F}
F_1(x) = -\frac{1}{8x} + o(|x|^{-1}) \;, \; {\rm when \;} x \to -\infty \;.
\end{eqnarray}
More generally, the analysis of Eq. (\ref{schrod_app}) suggests that $\tilde f(-\tilde r, x)$ admits an expansion of the form
\begin{eqnarray}\label{f_large_rcomplete}
\tilde f(-\tilde r, x) = \frac{1}{2^{7/6}} \tilde r^{-1/4} e^{-\frac{2}{3}\tilde r^{2/3} - x \sqrt{\tilde r}} \left(1 + \sum_{n=1}^\infty \frac{1}{\tilde r^{n/2}} F_n(x) \right)
\;, \; F_n(x) \underset{x \to -\infty}{\sim} \alpha_n  |x|^{-n} \;.
\end{eqnarray} 

Equipped with this asymptotic expansion (\ref{f_large_rcomplete}) we can now compute the large $\tilde r$ asymptotics of $\tilde p_{\rm typ}(\tilde r)$ beyond leading order. As this was done in the main text, the starting point of our analysis is the following expression
\begin{eqnarray}\label{start_gap_app}
\tilde p_{\rm typ}(\tilde r) = \frac{2^{1/3}}{\pi} \int_{-\infty}^\infty \left[\tilde f^2(-\tilde r,x) -\frac{1}{r^2} q^2(x) \tilde g^2(-\tilde r,x) \right] {\cal F}_2(x) \, dx \;.
\end{eqnarray}
We analyze the first term in (\ref{start_gap_app}) -- the integral involving $\tilde f^2(-\tilde r,x)$ -- by injecting the large $\tilde r$ expansion obtained above (\ref{f_large_rcomplete}). It yields
\begin{eqnarray}
 \frac{2^{1/3}}{\pi} \int_{-\infty}^\infty \tilde f^2(-\tilde r,x) = \frac{1}{4\pi} \frac{1}{\tilde r^{1/2}} e^{-\frac{4}{3} \tilde r^{3/2}} \int_{-\infty}^\infty e^{-2 x \sqrt{\tilde r}}\left(1 + \frac{2}{\sqrt{r}} F_1(x) + o(\tilde r^{-1/2})\right) {\cal F}_2(x) \, dx \;.
\end{eqnarray}
As we have seen before [see Eq. (\ref{saddle}) in the text], the integral over $x$ in (\ref{asympt_F2}) is dominated by the region where $x < 0$ where, as shown in the main text, one can use the saddle point method. Indeed, as shown previously in (\ref{saddle}), the saddle point is reached for $x^*= 2\sqrt{2} \, \tilde  r^{1/4}$. It is thus natural to perform the change of variable $x = u \tilde r^{1/4}$ and use the asymptotic behavior of $F_1(x)$ for large negative argument given in (\ref{asympt_F}) as well as the one for ${\cal F}_2(x)$ \cite{BBdF08}:
\begin{eqnarray}\label{asympt_F2}
{\cal F}_2(x) = \tau_2 |x|^{-1/8} e^{-\frac{1}{12} |x|^3} \left(1 + \frac{3}{2^6 |x|^3} + {\cal O}(|x|^{-6}) \right) \;, \; \tau_2 = 2^{1/24} e^{\zeta'(-1)} \;,
\end{eqnarray}
to obtain after straightforward (though tedious) manipulations the following expansion:
\begin{eqnarray}\label{asympt_first_part}
\frac{2^{1/3}}{\pi} \int_{-\infty}^\infty \tilde f^2(-\tilde r,x) = A  \; \exp{\left(-\dfrac{4}{3}\tilde r^{3/2} + \dfrac{8}{3}\sqrt{2}\tilde r^{3/4}\right)}{\tilde r^{-{21}/{32}}} \left(1 + \dfrac{131 \sqrt{2}}{1536} \tilde r^{-3/4} + {\cal O}(\tilde r^{-3/2})\right) \;,
\end{eqnarray} 
with $A = 2^{-91/48} e^{\zeta'(-1)}/\sqrt{\pi}$.

We now analyze the second term in Eq. (\ref{start_gap_app}), which involves $\tilde g(-\tilde r,x)$ that does not contribute to leading order when $\tilde r \to \infty$. To analyze this term, it is sufficient to expand $\tilde g(-\tilde r,x)$ using Eq. (\ref{fgtilde_larger_gap}) as well as ${\cal F}_2(x)$ using Eq. (\ref{asympt_F2}) to leading order. One can then perform a large $\tilde r$ expansion of this term using again the saddle point method, as shown in the main text (\ref{saddle}). One obtains, after some manipulations:
\begin{equation}\label{asympt_second_part}
 -\frac{2^{1/3}}{\pi \tilde r^2} \int_{-\infty}^\infty \left[  q^2(x) \tilde g^2(-\tilde r,x) \right] {\cal F}_2(x) \, dx = A  \; \exp{\left(-\dfrac{4}{3}\tilde r^{3/2} + \dfrac{8}{3}\sqrt{2}\tilde r^{3/4}\right)}{\tilde r^{-{21}/{32}}} \times \left(-\sqrt{2} \tilde r^{-3/4} + {\cal O}(\tilde r^{-3/2})\right) 
\end{equation}   
Finally, combining these asymptotic expansions (\ref{asympt_first_part}) and (\ref{asympt_second_part}) one obtains from (\ref{start_gap_app})
\begin{eqnarray}\label{final_gap_app}
\tilde p_{\rm typ}(\tilde r) = A  \; \exp{\left(-\dfrac{4}{3}r^{3/2} + \dfrac{8}{3}\sqrt{2}r^{3/4}\right)}{r^{-{21}/{32}}} \left(1 - \dfrac{1405 \sqrt{2}}{1536} r^{-3/4} + {\cal O}(r^{-3/2})\right) \;,
\end{eqnarray} 
as announced in the text in Eq.~(\ref{final_gap}).

\section{The formula for the first gap, $\tilde{p}_{\rm typ}(\tilde{r})$, obtained by Witte, Bornemann and Forrester~(WBF)}\label{formula_WBF}

In Ref. \cite{WBF13}, WBF obtained a formula for $\tilde p_{\rm typ}(\tilde r)$ in terms of the components of a solution of a particular isomonodromic problem relating to the Painlev\'e II equation. In this paper, they used a method which differs significantly for the one developed in the present paper. The starting point of their computation \cite{WBF13} is a formula, obtained in Ref. \cite{FW07}, for the joint PDF of the first and second smallest eigenvalues at the hard edge of unitary ensembles. They could then use a transformation which relates the hard edge to the soft edge \cite{BF03} to obtain an expression for $\tilde{p}_{\rm typ}(\tilde{r})$ in terms of the Hamiltonian system for Painlev\'e II, denoted by $\{q,p;t, H\}$. We introduce the parameters $\alpha$ and $\alpha_1$ satisfying the relations $\alpha = \alpha_1 -1/2$. In the present case, $\alpha_1 = 2$, implying $\alpha = 3/2$. The Painlev\'e II Hamiltonian is
\begin{eqnarray}\label{hamiltonian}
H = - \frac{1}{2} \left(2q_\alpha^2 - p_\alpha + t \right)p_\alpha - \alpha_1 q_\alpha \;,
\end{eqnarray}  
while the corresponding equations of motion are
\begin{eqnarray}
\partial_t q_\alpha = \partial_{p_\alpha} H = p_\alpha - q_\alpha^2 - \frac{1}{2}t \;, \; \partial_t p_\alpha = \partial_{q_\alpha} H = 2 q_\alpha p_\alpha + \alpha_1 \;,
\end{eqnarray}
from which $q_\alpha(t)$ satisfies the Painlev\'e II equation with parameter $\alpha$
\begin{eqnarray}
\partial^2_t q_\alpha = 2 q_\alpha^3 + t q_\alpha + \alpha \;.
\end{eqnarray}
Note that $H$ satisfies the second-order second degree differential equation of Jimbo-Miwa-Okamoto $\sigma$ form for PII with $\alpha_1  = 2$, $\alpha=\frac32$:
\begin{eqnarray}
\ddot{H}^2+4\dot{H}^3+2\dot{H}[t\dot{H}-H]=1 \;.
\end{eqnarray}

In terms of the Hamiltonian system (\ref{hamiltonian}) (with $\alpha_1 = 2$), the expression of $\tilde p_{\rm typ}(\tilde r)$ obtained in Ref.~\cite{WBF13}  reads 
\begin{eqnarray}\label{WBF_1}
\tilde{p}_{\rm typ}(\tilde{r})= \int_{-\infty}^{\infty}  {\rm p}_{(2)}^{\rm soft}(t,t-\tilde{r}) \, dt \;,
\end{eqnarray}
with
\begin{eqnarray}\label{WBF_2}
{\rm p}_{(2)}^{\rm soft}(t,t-\tilde r)=\frac{t^{-5/2}}{4\pi}{\rm p}_{(1)}^{\rm soft}(t)\exp{\left(-\frac43 t^{3/2}\right)}
\exp{\left(\int_{2^{1/3}t}^\infty {\rm d}y \left\{(2q_{3/2}+\frac{4}{p_{3/2}})(-y)-\sqrt{2y}-\frac{5}{2y} \right\}\right)}\nonumber\\
\times ({\rm U}\partial_x {\rm V}-{\rm V} \partial_x {\rm U}) (-2^{1/3} \tilde r; -2^{1/3} t) \;.
\end{eqnarray}
In Eq. (\ref{WBF_2}), ${\rm p}_{(1)}^{\rm soft}(t)$ is given by \cite{WBF13}
\begin{eqnarray}
{\rm p}_{(1)}^{\rm soft}(t)={\rm K}^{\rm soft}(t,t) \exp{\left(-\int_s^\infty{\rm d}t \left(\sigma_{\rm II}(t)-\frac{\rm d}{{\rm d}t} {\rm log}\,{\rm K}^{\rm soft}(t,t) \right) \right)} \;,
\end{eqnarray}
where ${\rm K}^{\rm soft}(x,y)$ is the Airy kernel
\begin{eqnarray}
{\rm K}^{\rm soft}(x,y)= \frac{{\rm Ai}(x){\rm Ai}'(y)-{\rm Ai}(y){\rm Ai}'(x)}{x-y}
\end{eqnarray}
and
\begin{eqnarray}
\sigma_{\rm II}(t)=-2^{1/3} H(-2^3 t)  \;,
\end{eqnarray}
where the Hamiltonian $H$ is given in Eq. (\ref{hamiltonian}). Finally, in Eq. (\ref{WBF_2}), ${\rm U} \equiv {\rm U}(x,t)$ and ${\rm V} = {\rm V}(x,t)$ are solutions of the following Lax pair:
\begin{eqnarray}
\partial_x \begin{pmatrix}{\rm U} \\{\rm V}\end{pmatrix} &=&\left\{
\begin{pmatrix}0&0 \\-\dfrac12&0\end{pmatrix} x +\begin{pmatrix}-q_{3/2}-\dfrac{2}{p_{3/2}}&-1 \\\dfrac12(t-p_{3/2})+[q_{3/2}+\dfrac{2}{p_{3/2}}]^2&q_{3/2}+\dfrac{2}{p_{3/2}}\end{pmatrix}+\begin{pmatrix}1&{p_{3/2}} \\0&-1\end{pmatrix}\frac1x
\right\}\begin{pmatrix}{\rm U} \\{\rm V}\end{pmatrix},\\
\partial_t \begin{pmatrix}{\rm U} \\{\rm V}\end{pmatrix} &=&\left\{
\begin{pmatrix}0&0 \\\dfrac12&0\end{pmatrix} x +\begin{pmatrix}0&1 \\0&-2[q_{3/2}+\dfrac{2}{p_{3/2}}]\end{pmatrix}
\right\}\begin{pmatrix}{\rm U} \\{\rm V}\end{pmatrix} \;.\\
\end{eqnarray}
which satisfy the boundary conditions for $t,x, t-x \to -\infty$
\begin{eqnarray}
{\rm U}(x,t)&\sim& -t \frac{{\rm Ai}(2^{-1/3}(x-t))}{{\rm Ai}(-2^{-1/3}t)}\\
{\rm V}(x,t)&\sim& t \frac{2^{-1/3} {\rm Ai}'(2^{-1/3}(x-t))+\sqrt{-\frac{t}{2}}{\rm Ai}(2^{-1/3}(x-t))}{{\rm Ai}(-2^{-1/3}t)} \;.
\end{eqnarray}

\end{document}